\documentclass[traditabstract, rnote]{aa} 

\usepackage[varg]{txfonts}
\usepackage{natbib}
\usepackage{longtable}
\usepackage{graphicx}
\usepackage{hyperref}

\def\de{\mathrm{e}}
\def\dr{\mathrm{d}}
\def\ne{n_\mathrm{e}}
\def\ce{C_{ij}^\mathrm{e}}
\def\upse{\Upsilon_{ij}^{\mathrm{e}}}
\def\ryd{E_{\mathrm{H}}^\infty}
\def\me{m_{\mathrm{e}}}

\begin{document}

 \title{Effective collision strengths between \ion{Mg}{I} and electrons\thanks{Quantum mechanical calculations from Zatsarinny et al. 2009, Phys. Rev. A., 79, 052709.}}
 \author{T. Merle\inst{1} \and F. Th\'{e}venin\inst{2} \and O. Zatsarinny\inst{3}}
 \institute{Institut d'Astronomie et d'Astrophysique, Universit\'{e} Libre de Bruxelles, CP 226, Boulevard du Triomphe, 1050 Brussels, Belgium \and
            Universit\'{e} de Nice Sophia-Antipolis, Laboratoire Lagrange, UMR7293 CNRS, OCA, CS34229, F06304, Nice Cedex 4 \and
			Department of Physics and Astronomy, Drake University, Des Moines, IA 50311, USA
			}
 \date{Received ?; accepted ?}

 \abstract{The treatment of the inelastic collisions with electrons and hydrogen atoms are the main source of uncertainties in non-Local Thermodynamic Equilibrium (LTE) spectral line computations. We report, in this research note, quantum mechanical data for 369 collisional transitions of \ion{Mg}{I} with electrons for temperatures comprised between 500 and 20000~K. We give the quantum mechanical data in terms of effective collision strengths, more practical for non-LTE studies. 
 }

 
 \maketitle

 \section{Introduction}
Atomic data are the main ingredient for a good non-local thermodynamic equilibrium (non-LTE) description of the interaction of a chemical species with matter and radiation in atmospheres of late-type stars. Huge efforts have been made over the years regarding experimental and theoretical data (see e.g. \citealt{henry93}). Semi-classical or empirical formul\ae\ are widely used because they provide rough estimates of collision rates based on oscillator strength for radiatively permitted transitions while no formul\ae\ exist to correctly treat the forbidden transitions. The most used of such formul\ae\ is the impact parameter method (IPM, \citealt{seaton62a,seaton62b}) which implies computing collision cross-sections for weak and strong coupling. A simpler way (but also not as precise) is to use the semi-empirical formulation from \citet{van_regemorter62} which is an approximation of the IPM using experimental data available at that time. A more recent version that takes into account recent experimental data give a formulation with a dependance on the change of the principal quantum number \citep{fisher96} but this version is not used very often in the literature.

Magnesium is an $\alpha$-element produced by type~II supernov\ae\ of which is important to study its enrichment relative to iron in different stellar populations. The numerous optical lines of \ion{Mg}{i}, including the green b triplet, allow the determination of its abundance in many kinds of stars which is why a good description of its atomic data and especially the inelastic collisions are needed. \citet{zatsarinny09} lead a very detailed study of angle-differential cross-sections for electron scattering from neutral magnesium using a B-spline R-matrix (BSR) method. However, only results for five collision cross-sections with the the ground state were published in this paper, whereas many collision cross-sections were computed. In the present research note, we decided to publish the other data in the form of effective collision strengths useful for non-LTE applications and which concern 369 transitions between excited states.

\section{Previous calculations}
\begin{figure}[h!]
   \includegraphics[width=\linewidth]{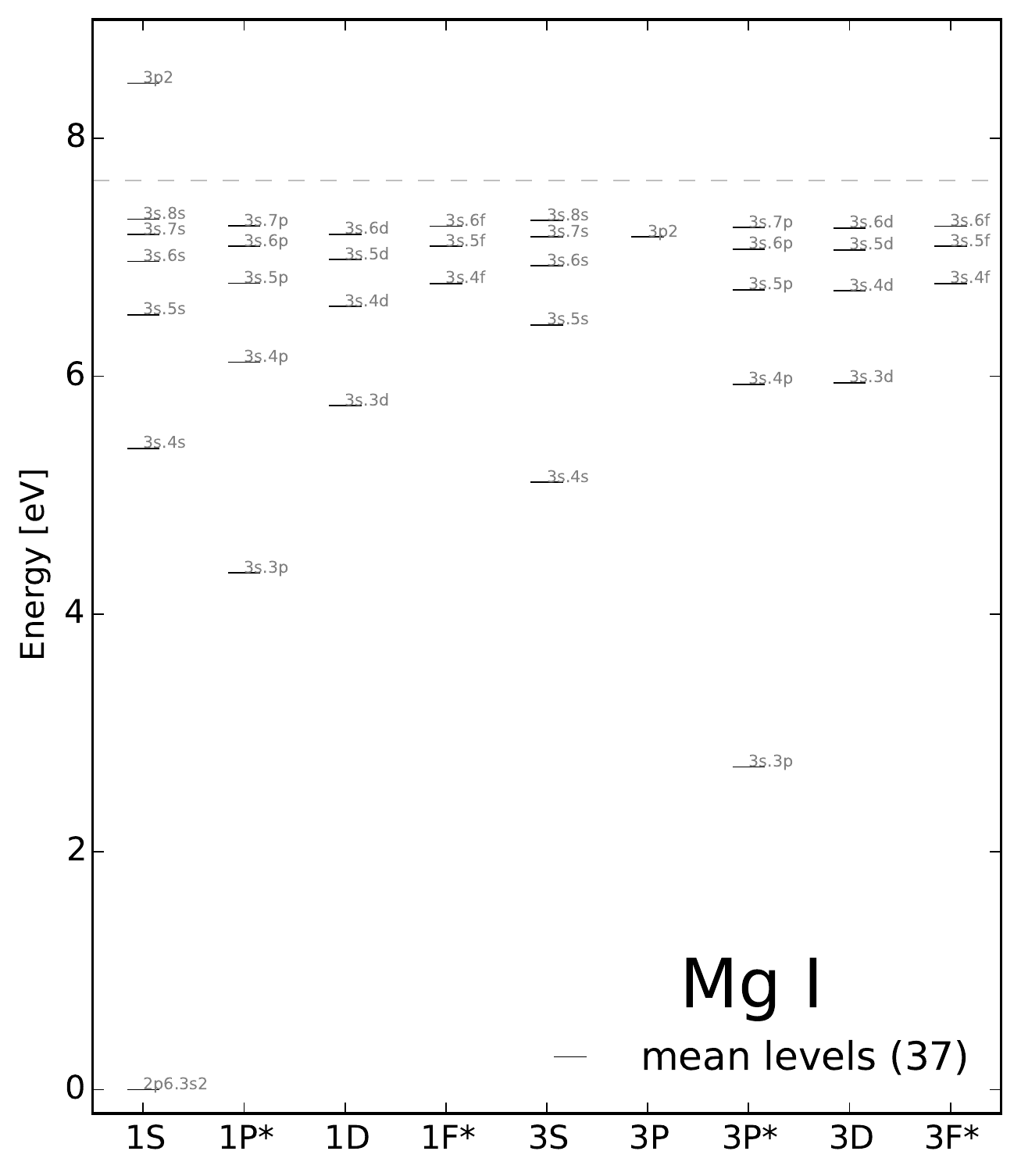}

     \includegraphics[width=\linewidth]{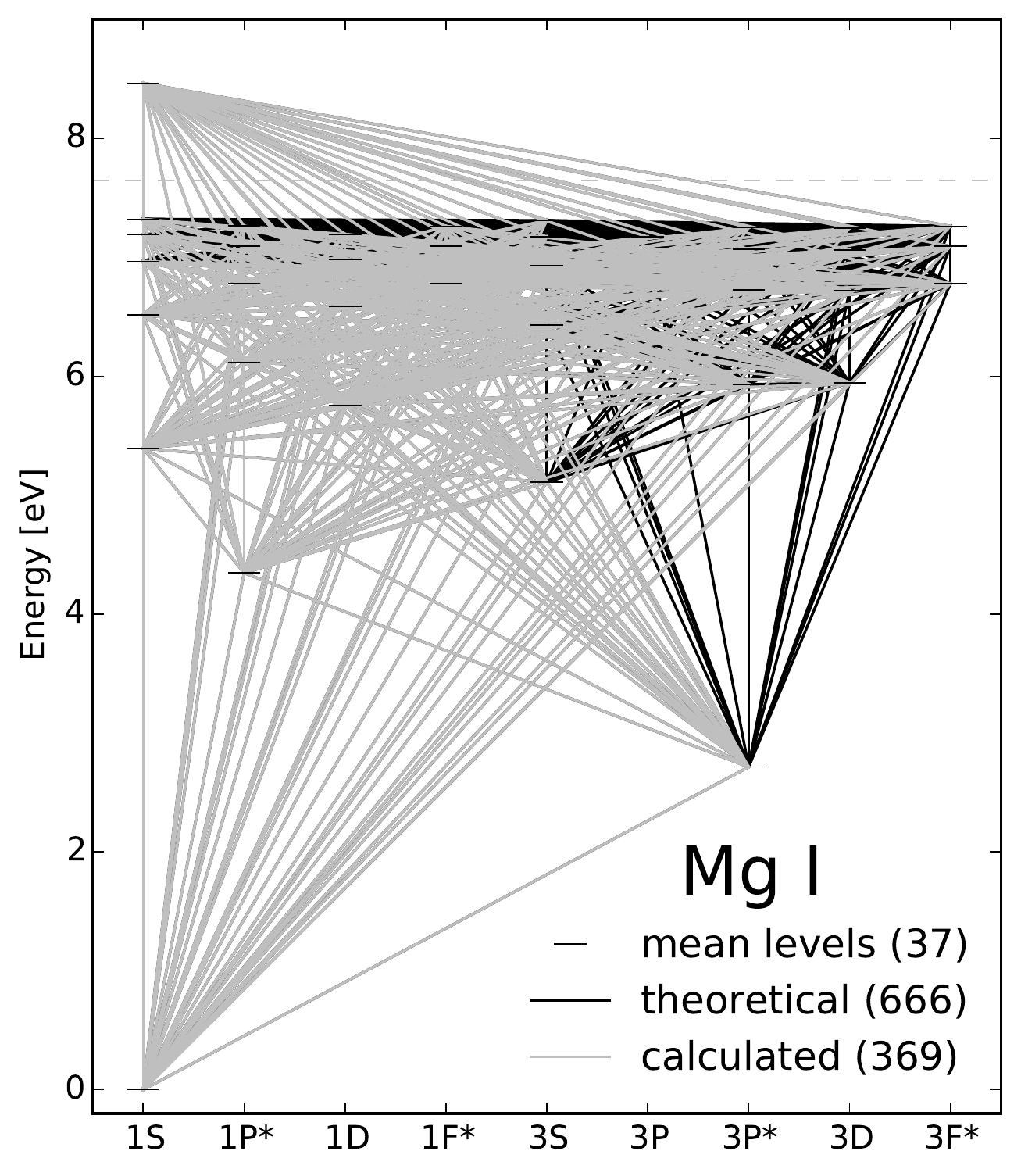}
       \caption{Grotrian diagram of \ion{Mg}{I}  with mean energy levels labeled (top) and with the collision transitions (bottom). Theoretical collision transitions are shown with black lines whereas the calculated ones are shown with grey lines.}
         \label{fig:grot_mgi}
	  \end{figure}

In \citet{zatsarinny09}, 37 target states were included in the scattering calculations. The configurations, terms, energy levels, and statistical weights $g$ are given in Table~\ref{tab:ael}. The Grotrian diagram of the energy levels are given in the top panel of Fig.~\ref{fig:grot_mgi}. The double excited level 3p$^2$ $^1$S located above the ionization energy is also included owing to big mixing of this configuration to the ground state. This improves the convergence of the total scattering function. The quality of the model atom description can be assessed from the theoretical oscillator strengths they calculated compared with the NIST\footnote{Available at: \url{http://www.nist.gov/pml/data/asd.cfm}} counterparts. The differences are less than 3~\%. Accurate oscillator strengths are important for subsequent calculations of collision cross-sections. 

The B-spline R-matrix code \citep{zatsarinny06} was used for the scattering calculations. Angle-differential cross-sections for electron-impact excitation are obtained and described in detail in \citet{zatsarinny09}. In their Fig.~8, they show angle-integrated cross-sections for five transitions with the ground state but cross-sections for 369 transitions were actually obtained and not published. Such cross-sections are extremely important for non-LTE studies. With 37 levels included, there are a total of 666 collision transition possibilities. We present here all the transitions between the first 17 levels up to 3s5p~$^1$P$^\circ$, and all effective collision strengths for excitation of these levels to the more excited level indicated in Table~\ref{tab:ael}.There are 369 overall (see bottom panel of Fig.~\ref{fig:grot_mgi}). Though the cross-section for transitions between highly-excited levels can also be extracted from the BSR calculation, they cannot be considered reliable because of slow partial wave convergence for these levels.  
In non-LTE calculations, we hardly need detailed cross-sections, but rather thermal average quantities such as effective collision strengths.

 \begin{table}
   \center
     \caption{Energy levels of \ion{Mg}{I}.}
      \begin{tabular}{rllcrr}
       Index & Conf. & Term & E [eV]  & E [cm$^{-1}$] & g \\
        \hline
	 1 & 3s$^2$ & $^1$S         & 0.000 &     0.0000 &  1 \\
	 2 & 3s3p   & $^3$P$^\circ$ & 2.645 & 21890.8541 &  9 \\
	 3 & 3s3p   & $^1$P$^\circ$ & 4.286 & 35051.2640 &  3 \\
	 4 & 3s4s   & $^3$S         & 5.015 & 41197.4030 &  3 \\
	 5 & 3s4s   & $^1$S         & 5.301 & 43503.3330 &  1 \\
	 6 & 3s3d   & $^1$D         & 5.640 & 46403.0650 &  5 \\
	 7 & 3s4p   & $^3$P$^\circ$ & 5.822 & 47847.7968 &  9 \\
	 8 & 3s3d   & $^3$D         & 5.834 & 47957.0416 & 15 \\
	 9 & 3s4p   & $^1$P$^\circ$ & 6.018 & 49346.7290 &  3 \\
	10 & 3s5s   & $^3$S         & 6.321 & 51872.5260 &  3 \\
	11 & 3s5s   & $^1$S         & 6.407 & 52556.2060 &  1 \\
	12 & 3s4d   & $^1$D         & 6.473 & 53134.6420 &  5 \\
	13 & 3s4d   & $^3$D         & 6.603 & 54192.2845 & 15 \\
	14 & 3s5p   & $^3$P$^\circ$ & 6.611 & 54251.4108 &  9 \\
	15 & 3s4f   & $^1$F$^\circ$ & 6.660 & 54676.4380 &  7 \\
	16 & 3s4f   & $^3$F$^\circ$ & 6.660 & 54676.7130 & 21 \\
	17 & 3s5p   & $^1$P$^\circ$ & 6.671 & 54706.5360 &  3 \\
	18 & 3s6s   & $^3$S         & 6.815 & 55891.8000 &  3 \\
	19 & 3s6s   & $^1$S         & 6.852 & 56186.8730 &  1 \\
	20 & 3s5d   & $^1$D         & 6.865 & 56308.3810 &  5 \\
	21 & 3s5d   & $^3$D         & 6.946 & 56968.2386 & 15 \\
	22 & 3s6p   & $^3$P$^\circ$ & 6.952 & 57018.3750 &  9 \\
	23 & 3s5f   & $^1$F$^\circ$ & 6.973 & 57204.1630 &  7 \\
	24 & 3s5f   & $^3$F$^\circ$ & 6.973 & 57204.2740 & 21 \\
	25 & 3s6p   & $^1$P$^\circ$ & 6.978 & 57214.9920 &  3 \\
	26 & 3s7s   & $^3$S         & 7.057 & 57855.2140 &  3 \\
	27 & 3s6d   & $^1$D         & 7.078 & 58023.2460 &  5 \\
	28 & 3s7s   & $^1$S         & 7.078 & 58009.4100 &  1 \\
	29 & 3p$^2$ & $^3$P         & 7.081 & 57853.6300 &  9 \\
	30 & 3s6d   & $^3$D         & 7.133 & 58442.8469 & 15 \\
	31 & 3s7p   & $^3$P$^\circ$ & 7.140 & 58477.3943 &  9 \\
	32 & 3s6f   & $^3$F$^\circ$ & 7.148 & 58575.5270 & 21 \\
  33 & 3s6f   & $^1$F$^\circ$ & 7.148 & 58575.4770 &  7 \\
  34 & 3s7p   & $^1$P$^\circ$ & 7.156 & 58580.2300 &  3 \\
  35 & 3s8s   & $^3$S         & 7.225 & 58962.7390 &  3 \\
  36 & 3s8s   & $^1$S         & 7.245 & 59053.5200 &  1 \\
  37 & 3p$^2$ & $^1$S         & 8.476 & 68275.0000 &  1 \\
\hline
\end{tabular}
  \label{tab:ael}
   \end{table}

 \section{Effective collision strengths}

The collision rate from state $i$ to state $j$ with electrons $\ce$ can be written as 
\begin{equation}
\ce(T)=\ne~\mathcal{A} ~\frac{\upse(T)}{g_i\sqrt{T}} \de^{-E_{ij}/kT} ~~~[\mathrm{s^{-1}}]
\end{equation}
where $\ne$ is the electron density, $g_i$ the statistical weight of the lower level, $E_{ij}$ the transition energy, $k$ the Boltzmann constant, $T$ the temperature of the medium and $\mathcal{A}$ the constant
\begin{equation}
\mathcal{A}=\pi a_0^2  \left[\frac{8\ryd}{\pi \me}\right]^{1/2}\left[\frac{\ryd}{k}\right]^{1/2} = 8.629 \times 10^{-6} \mathrm{[cm^3~s^{-1}~K^{1/2}],}
\end{equation}
where $a_0$ is the Bohr radius, $\ryd$ is the Rydberg unit of energy, and $\me$ the mass of electron. Moreover, we use the effective collision strenght $\upse$ defined as
\begin{equation}
\label{eq:upsilon}
\upse = g_i \frac{kT}{\ryd} \int_0^\infty \sigma_{ij}(x) (x+x_{ij})\de^{-x} \dr x
\end{equation}
where $x=E/kT$ and $x_{ij}=E_{ij}/kT$ are the kinetic energy after excitation and the energy of the transition in unit of $kT$, $\sigma_{ij}$ is the collision cross-section expressed in unit of $\pi a_0^2$, and $\upse$ is dimensionless and symmetric with respect to the transition (i.e. $\upse=\Upsilon_{ji}^\mathrm{e}$). The resulting effective collision strengths are given in Table~\ref{tab:ups}\footnote{Table~\ref{tab:ups} available electronically only.} for nine temperatures between 500 and 20000~K. The level indexes in Table~\ref{tab:ups} refer to thr levels in Table~\ref{tab:ael}.

These data will be useful for non-LTE studies of magnesium in stellar atmospheres of cool stars since they come from reliable quantum mechanical calculations. We should mention that such efforts toward calculating and collecting quantum mechanical data for inelastic collisions is extremely important so as to derive reliable non-LTE abundance corrections. Special efforts are also currently underway for computing inelastic collisions with hydrogen atoms \citep{barklem11, belyaev14}.

 \begin{acknowledgements}
TM is short-term foreign postdoctoral fellow from FNRS (FRFC convention, ref: PDR T.0198.13).
 \end{acknowledgements}

 \bibliography{biblio}{} 
 \bibliographystyle{aa} 

 \onecolumn
  \tiny
  \begin{longtable}{rrccccccccc}
   \caption{Effective collision strengths. \label{tab:ups}}\\
    \multicolumn{2}{c}{Level index} & \multicolumn{9}{c}{Temperature [K]}\\
    \cline{3-11}
    $i$ & $j$ & 500  & 1000 & 2000 & 4000 & 6000 & 8000 & 10000 & 15000 & 20000 \\
   \hline\\
   \endfirsthead
   \caption{Continued.}\\
       \multicolumn{2}{c}{Level index} & \multicolumn{9}{c}{Temperature [K]}\\
    \cline{3-11}
    $i$ & $j$ & 500  & 1000 & 2000 & 4000 & 6000 & 8000 & 10000 & 15000 & 20000 \\ 
    \hline\\
   \endhead
   \hline
   \endfoot
  1 &  2 & 3.14e$-02$ & 9.78e$-02$ & 3.09e$-01$ & 9.00e$-01$ & 1.58e$+00$ & 2.28e$+00$ & 2.97e$+00$ & 4.59e$+00$ & 6.04e$+00$ \\
  1 &  3 & 1.63e$-02$ & 3.74e$-02$ & 9.64e$-02$ & 2.96e$-01$ & 5.98e$-01$ & 9.92e$-01$ & 1.47e$+00$ & 3.06e$+00$ & 5.21e$+00$ \\
  1 &  4 & 2.04e$-03$ & 6.55e$-03$ & 2.05e$-02$ & 5.69e$-02$ & 9.56e$-02$ & 1.33e$-01$ & 1.70e$-01$ & 2.54e$-01$ & 3.30e$-01$ \\
  1 &  5 & 6.92e$-04$ & 3.81e$-03$ & 1.62e$-02$ & 4.93e$-02$ & 8.54e$-02$ & 1.23e$-01$ & 1.62e$-01$ & 2.69e$-01$ & 3.91e$-01$ \\
  1 &  6 & 4.23e$-03$ & 1.20e$-02$ & 3.03e$-02$ & 7.26e$-02$ & 1.24e$-01$ & 1.85e$-01$ & 2.58e$-01$ & 4.95e$-01$ & 8.18e$-01$ \\
  1 &  7 & 8.56e$-03$ & 1.91e$-02$ & 4.15e$-02$ & 8.85e$-02$ & 1.35e$-01$ & 1.80e$-01$ & 2.23e$-01$ & 3.27e$-01$ & 4.27e$-01$ \\
  1 &  8 & 2.39e$-03$ & 6.95e$-03$ & 2.23e$-02$ & 6.82e$-02$ & 1.26e$-01$ & 1.92e$-01$ & 2.64e$-01$ & 4.67e$-01$ & 6.94e$-01$ \\
  1 &  9 & 2.72e$-03$ & 7.48e$-03$ & 2.09e$-02$ & 5.46e$-02$ & 9.59e$-02$ & 1.46e$-01$ & 2.08e$-01$ & 4.18e$-01$ & 7.16e$-01$ \\
  1 & 10 & 1.66e$-03$ & 4.70e$-03$ & 1.10e$-02$ & 2.23e$-02$ & 3.27e$-02$ & 4.27e$-02$ & 5.27e$-02$ & 7.82e$-02$ & 1.04e$-01$ \\
  1 & 11 & 1.89e$-03$ & 4.42e$-03$ & 9.35e$-03$ & 1.91e$-02$ & 2.95e$-02$ & 4.12e$-02$ & 5.45e$-02$ & 9.52e$-02$ & 1.46e$-01$ \\
  1 & 12 & 3.32e$-03$ & 7.56e$-03$ & 1.60e$-02$ & 3.25e$-02$ & 5.02e$-02$ & 7.03e$-02$ & 9.37e$-02$ & 1.66e$-01$ & 2.57e$-01$ \\
  1 & 13 & 2.94e$-03$ & 6.95e$-03$ & 1.55e$-02$ & 3.47e$-02$ & 5.72e$-02$ & 8.35e$-02$ & 1.14e$-01$ & 2.05e$-01$ & 3.12e$-01$ \\
  1 & 14 & 1.73e$-03$ & 4.00e$-03$ & 9.18e$-03$ & 2.09e$-02$ & 3.33e$-02$ & 4.62e$-02$ & 5.97e$-02$ & 9.55e$-02$ & 1.33e$-01$ \\
  1 & 15 & 1.59e$-03$ & 3.39e$-03$ & 6.70e$-03$ & 1.29e$-02$ & 1.89e$-02$ & 2.51e$-02$ & 3.14e$-02$ & 4.82e$-02$ & 6.65e$-02$ \\
  1 & 16 & 9.44e$-04$ & 2.49e$-03$ & 5.49e$-03$ & 1.12e$-02$ & 1.75e$-02$ & 2.45e$-02$ & 3.22e$-02$ & 5.29e$-02$ & 7.38e$-02$ \\
  1 & 17 & 1.16e$-03$ & 3.21e$-03$ & 8.00e$-03$ & 1.92e$-02$ & 3.29e$-02$ & 4.99e$-02$ & 7.10e$-02$ & 1.43e$-01$ & 2.43e$-01$ \\
  1 & 18 & 9.68e$-04$ & 2.06e$-03$ & 4.01e$-03$ & 7.79e$-03$ & 1.16e$-02$ & 1.57e$-02$ & 2.00e$-02$ & 3.16e$-02$ & 4.37e$-02$ \\
  1 & 19 & 6.46e$-04$ & 1.42e$-03$ & 3.05e$-03$ & 6.74e$-03$ & 1.11e$-02$ & 1.65e$-02$ & 2.28e$-02$ & 4.26e$-02$ & 6.68e$-02$ \\
  1 & 20 & 1.96e$-03$ & 4.51e$-03$ & 9.09e$-03$ & 1.78e$-02$ & 2.81e$-02$ & 4.05e$-02$ & 5.48e$-02$ & 9.64e$-02$ & 1.44e$-01$ \\
  1 & 21 & 1.30e$-03$ & 2.95e$-03$ & 6.38e$-03$ & 1.60e$-02$ & 3.07e$-02$ & 5.01e$-02$ & 7.31e$-02$ & 1.41e$-01$ & 2.18e$-01$ \\
  1 & 22 & 9.44e$-04$ & 2.31e$-03$ & 5.12e$-03$ & 1.15e$-02$ & 1.94e$-02$ & 2.90e$-02$ & 3.98e$-02$ & 6.99e$-02$ & 1.02e$-01$ \\
  1 & 23 & 1.17e$-03$ & 2.50e$-03$ & 5.05e$-03$ & 1.01e$-02$ & 1.59e$-02$ & 2.24e$-02$ & 2.95e$-02$ & 4.89e$-02$ & 6.93e$-02$ \\
  1 & 24 & 5.97e$-04$ & 1.42e$-03$ & 2.99e$-03$ & 6.69e$-03$ & 1.16e$-02$ & 1.74e$-02$ & 2.38e$-02$ & 4.08e$-02$ & 5.74e$-02$ \\
  1 & 25 & 7.72e$-04$ & 1.95e$-03$ & 4.40e$-03$ & 9.28e$-03$ & 1.47e$-02$ & 2.12e$-02$ & 2.92e$-02$ & 5.68e$-02$ & 9.54e$-02$ \\
  1 & 26 & 5.60e$-04$ & 1.22e$-03$ & 2.57e$-03$ & 5.00e$-03$ & 7.26e$-03$ & 9.53e$-03$ & 1.18e$-02$ & 1.76e$-02$ & 2.33e$-02$ \\
  1 & 27 & 1.68e$-03$ & 4.43e$-03$ & 1.17e$-02$ & 3.00e$-02$ & 5.01e$-02$ & 7.06e$-02$ & 9.12e$-02$ & 1.42e$-01$ & 1.91e$-01$ \\
  1 & 28 & 3.32e$-04$ & 8.13e$-04$ & 1.98e$-03$ & 4.54e$-03$ & 7.39e$-03$ & 1.07e$-02$ & 1.47e$-02$ & 2.66e$-02$ & 4.08e$-02$ \\
  1 & 29 & 6.47e$-06$ & 3.77e$-05$ & 2.27e$-04$ & 1.28e$-03$ & 3.20e$-03$ & 5.77e$-03$ & 8.81e$-03$ & 1.77e$-02$ & 2.77e$-02$ \\
  1 & 30 & 5.75e$-04$ & 1.65e$-03$ & 5.35e$-03$ & 1.88e$-02$ & 3.77e$-02$ & 5.98e$-02$ & 8.39e$-02$ & 1.48e$-01$ & 2.13e$-01$ \\
  1 & 31 & 6.79e$-04$ & 1.88e$-03$ & 5.63e$-03$ & 1.71e$-02$ & 3.25e$-02$ & 5.03e$-02$ & 6.91e$-02$ & 1.17e$-01$ & 1.61e$-01$ \\
  1 & 32 & 4.69e$-04$ & 1.16e$-03$ & 2.83e$-03$ & 7.06e$-03$ & 1.22e$-02$ & 1.79e$-02$ & 2.39e$-02$ & 3.91e$-02$ & 5.34e$-02$ \\
  1 & 33 & 4.44e$-04$ & 1.24e$-03$ & 3.77e$-03$ & 1.29e$-02$ & 2.63e$-02$ & 4.22e$-02$ & 5.95e$-02$ & 1.05e$-01$ & 1.52e$-01$ \\
  1 & 34 & 2.70e$-04$ & 6.98e$-04$ & 2.18e$-03$ & 8.36e$-03$ & 1.89e$-02$ & 3.32e$-02$ & 5.04e$-02$ & 1.04e$-01$ & 1.68e$-01$ \\
  1 & 35 & 1.14e$-04$ & 4.24e$-04$ & 2.05e$-03$ & 8.53e$-03$ & 1.64e$-02$ & 2.41e$-02$ & 3.13e$-02$ & 4.70e$-02$ & 6.07e$-02$ \\
  1 & 36 & 7.59e$-05$ & 1.97e$-04$ & 8.12e$-04$ & 4.11e$-03$ & 1.01e$-02$ & 1.84e$-02$ & 2.84e$-02$ & 5.85e$-02$ & 9.25e$-02$ \\
  1 & 37 & 5.77e$-05$ & 1.95e$-04$ & 6.33e$-04$ & 2.61e$-03$ & 6.54e$-03$ & 1.24e$-02$ & 1.98e$-02$ & 4.32e$-02$ & 7.12e$-02$ \\
  2 &  3 & 3.78e$-02$ & 8.45e$-02$ & 2.02e$-01$ & 5.43e$-01$ & 9.82e$-01$ & 1.47e$+00$ & 1.98e$+00$ & 3.26e$+00$ & 4.53e$+00$ \\
  2 &  4 & 4.88e$-04$ & 2.02e$-03$ & 8.64e$-03$ & 3.89e$-02$ & 9.75e$-02$ & 1.91e$-01$ & 3.24e$-01$ & 8.58e$-01$ & 1.70e$+00$ \\
  2 &  5 & 1.43e$-02$ & 3.87e$-02$ & 9.97e$-02$ & 2.32e$-01$ & 3.61e$-01$ & 4.82e$-01$ & 5.96e$-01$ & 8.53e$-01$ & 1.07e$+00$ \\
  2 &  6 & 2.55e$-02$ & 6.55e$-02$ & 1.56e$-01$ & 3.60e$-01$ & 5.91e$-01$ & 8.41e$-01$ & 1.11e$+00$ & 1.82e$+00$ & 2.60e$+00$ \\
  2 &  9 & 4.67e$-03$ & 1.71e$-02$ & 5.54e$-02$ & 1.48e$-01$ & 2.48e$-01$ & 3.53e$-01$ & 4.62e$-01$ & 7.44e$-01$ & 1.03e$+00$ \\
  2 & 11 & 8.44e$-03$ & 1.84e$-02$ & 3.90e$-02$ & 8.11e$-02$ & 1.23e$-01$ & 1.64e$-01$ & 2.05e$-01$ & 3.02e$-01$ & 3.93e$-01$ \\
  2 & 12 & 7.89e$-03$ & 1.76e$-02$ & 3.96e$-02$ & 9.14e$-02$ & 1.53e$-01$ & 2.25e$-01$ & 3.09e$-01$ & 5.61e$-01$ & 8.61e$-01$ \\
  2 & 15 & 1.65e$-03$ & 3.83e$-03$ & 8.99e$-03$ & 2.13e$-02$ & 3.61e$-02$ & 5.33e$-02$ & 7.26e$-02$ & 1.26e$-01$ & 1.82e$-01$ \\
  2 & 17 & 3.04e$-03$ & 8.84e$-03$ & 2.28e$-02$ & 5.60e$-02$ & 9.42e$-02$ & 1.36e$-01$ & 1.82e$-01$ & 3.03e$-01$ & 4.27e$-01$ \\
  2 & 19 & 2.58e$-03$ & 5.72e$-03$ & 1.28e$-02$ & 2.98e$-02$ & 4.89e$-02$ & 6.93e$-02$ & 9.05e$-02$ & 1.44e$-01$ & 1.97e$-01$ \\
  2 & 20 & 4.37e$-03$ & 1.02e$-02$ & 2.18e$-02$ & 4.74e$-02$ & 8.07e$-02$ & 1.23e$-01$ & 1.74e$-01$ & 3.29e$-01$ & 5.08e$-01$ \\
  2 & 23 & 1.33e$-03$ & 3.45e$-03$ & 7.95e$-03$ & 1.84e$-02$ & 3.15e$-02$ & 4.66e$-02$ & 6.29e$-02$ & 1.07e$-01$ & 1.51e$-01$ \\
  2 & 25 & 1.89e$-03$ & 4.78e$-03$ & 1.11e$-02$ & 2.54e$-02$ & 4.27e$-02$ & 6.29e$-02$ & 8.52e$-02$ & 1.45e$-01$ & 2.06e$-01$ \\
  2 & 27 & 3.07e$-03$ & 8.12e$-03$ & 2.18e$-02$ & 5.90e$-02$ & 1.05e$-01$ & 1.56e$-01$ & 2.11e$-01$ & 3.52e$-01$ & 4.95e$-01$ \\
  2 & 28 & 1.38e$-03$ & 3.20e$-03$ & 7.72e$-03$ & 1.78e$-02$ & 2.85e$-02$ & 4.01e$-02$ & 5.25e$-02$ & 8.50e$-02$ & 1.17e$-01$ \\
  2 & 33 & 1.23e$-03$ & 2.91e$-03$ & 7.61e$-03$ & 2.37e$-02$ & 4.64e$-02$ & 7.31e$-02$ & 1.02e$-01$ & 1.81e$-01$ & 2.60e$-01$ \\
  2 & 34 & 8.64e$-04$ & 2.22e$-03$ & 7.19e$-03$ & 2.76e$-02$ & 5.85e$-02$ & 9.53e$-02$ & 1.35e$-01$ & 2.35e$-01$ & 3.32e$-01$ \\
  2 & 36 & 3.27e$-04$ & 9.68e$-04$ & 3.88e$-03$ & 1.66e$-02$ & 3.57e$-02$ & 5.87e$-02$ & 8.37e$-02$ & 1.49e$-01$ & 2.13e$-01$ \\
  2 & 37 & 3.43e$-04$ & 1.35e$-03$ & 5.03e$-03$ & 1.85e$-02$ & 3.93e$-02$ & 6.62e$-02$ & 9.79e$-02$ & 1.91e$-01$ & 2.94e$-01$ \\
  3 &  4 & 3.69e$-02$ & 9.94e$-02$ & 2.63e$-01$ & 6.29e$-01$ & 9.79e$-01$ & 1.29e$+00$ & 1.58e$+00$ & 2.19e$+00$ & 2.70e$+00$ \\
  3 &  5 & 7.19e$-02$ & 2.22e$-01$ & 6.32e$-01$ & 1.64e$+00$ & 2.86e$+00$ & 4.33e$+00$ & 6.06e$+00$ & 1.15e$+01$ & 1.85e$+01$ \\
  3 &  6 & 7.52e$-02$ & 1.95e$-01$ & 4.73e$-01$ & 1.20e$+00$ & 2.15e$+00$ & 3.33e$+00$ & 4.73e$+00$ & 9.20e$+00$ & 1.50e$+01$ \\
  3 &  7 & 8.06e$-02$ & 1.89e$-01$ & 4.25e$-01$ & 9.23e$-01$ & 1.41e$+00$ & 1.88e$+00$ & 2.31e$+00$ & 3.27e$+00$ & 4.08e$+00$ \\
  3 &  8 & 2.78e$-02$ & 7.61e$-02$ & 2.11e$-01$ & 5.53e$-01$ & 9.31e$-01$ & 1.32e$+00$ & 1.71e$+00$ & 2.67e$+00$ & 3.63e$+00$ \\
  3 &  9 & 2.42e$-02$ & 6.92e$-02$ & 2.05e$-01$ & 5.77e$-01$ & 1.04e$+00$ & 1.58e$+00$ & 2.20e$+00$ & 4.14e$+00$ & 6.61e$+00$ \\
  3 & 10 & 6.62e$-03$ & 1.78e$-02$ & 4.37e$-02$ & 1.01e$-01$ & 1.59e$-01$ & 2.16e$-01$ & 2.70e$-01$ & 3.96e$-01$ & 5.09e$-01$ \\
  3 & 11 & 1.26e$-02$ & 2.73e$-02$ & 5.67e$-02$ & 1.16e$-01$ & 1.78e$-01$ & 2.48e$-01$ & 3.30e$-01$ & 5.93e$-01$ & 9.46e$-01$ \\
  3 & 12 & 2.42e$-02$ & 6.16e$-02$ & 1.48e$-01$ & 3.44e$-01$ & 5.60e$-01$ & 7.94e$-01$ & 1.05e$+00$ & 1.74e$+00$ & 2.53e$+00$ \\
  3 & 13 & 1.83e$-02$ & 4.43e$-02$ & 1.03e$-01$ & 2.23e$-01$ & 3.40e$-01$ & 4.56e$-01$ & 5.73e$-01$ & 8.71e$-01$ & 1.17e$+00$ \\
  3 & 14 & 9.96e$-03$ & 2.48e$-02$ & 5.82e$-02$ & 1.32e$-01$ & 2.09e$-01$ & 2.86e$-01$ & 3.62e$-01$ & 5.48e$-01$ & 7.27e$-01$ \\
  3 & 15 & 9.85e$-03$ & 3.01e$-02$ & 8.42e$-02$ & 2.30e$-01$ & 4.26e$-01$ & 6.74e$-01$ & 9.75e$-01$ & 1.94e$+00$ & 3.17e$+00$ \\
  3 & 16 & 1.06e$-02$ & 2.14e$-02$ & 4.35e$-02$ & 8.92e$-02$ & 1.41e$-01$ & 2.00e$-01$ & 2.66e$-01$ & 4.48e$-01$ & 6.33e$-01$ \\
  3 & 17 & 1.06e$-02$ & 2.96e$-02$ & 7.25e$-02$ & 1.64e$-01$ & 2.68e$-01$ & 3.91e$-01$ & 5.34e$-01$ & 9.89e$-01$ & 1.57e$+00$ \\
  3 & 18 & 2.68e$-03$ & 7.25e$-03$ & 1.71e$-02$ & 3.87e$-02$ & 6.11e$-02$ & 8.33e$-02$ & 1.05e$-01$ & 1.55e$-01$ & 2.00e$-01$ \\
  3 & 19 & 4.21e$-03$ & 1.03e$-02$ & 2.26e$-02$ & 4.64e$-02$ & 7.20e$-02$ & 1.02e$-01$ & 1.38e$-01$ & 2.53e$-01$ & 4.02e$-01$ \\
  3 & 20 & 1.16e$-02$ & 2.83e$-02$ & 6.32e$-02$ & 1.46e$-01$ & 2.52e$-01$ & 3.79e$-01$ & 5.24e$-01$ & 9.65e$-01$ & 1.51e$+00$ \\
  3 & 21 & 6.81e$-03$ & 1.80e$-02$ & 4.36e$-02$ & 1.04e$-01$ & 1.75e$-01$ & 2.53e$-01$ & 3.35e$-01$ & 5.44e$-01$ & 7.49e$-01$ \\
  3 & 22 & 5.03e$-03$ & 1.32e$-02$ & 3.04e$-02$ & 6.81e$-02$ & 1.14e$-01$ & 1.67e$-01$ & 2.25e$-01$ & 3.80e$-01$ & 5.35e$-01$ \\
  3 & 23 & 8.71e$-03$ & 2.32e$-02$ & 5.44e$-02$ & 1.35e$-01$ & 2.52e$-01$ & 4.10e$-01$ & 6.07e$-01$ & 1.24e$+00$ & 2.03e$+00$ \\
  3 & 24 & 3.64e$-03$ & 1.02e$-02$ & 2.52e$-02$ & 5.99e$-02$ & 1.02e$-01$ & 1.50e$-01$ & 2.00e$-01$ & 3.28e$-01$ & 4.48e$-01$ \\
  3 & 25 & 7.00e$-03$ & 1.78e$-02$ & 4.00e$-02$ & 8.69e$-02$ & 1.45e$-01$ & 2.16e$-01$ & 2.99e$-01$ & 5.50e$-01$ & 8.47e$-01$ \\
  3 & 26 & 1.99e$-03$ & 4.96e$-03$ & 1.15e$-02$ & 2.38e$-02$ & 3.48e$-02$ & 4.50e$-02$ & 5.46e$-02$ & 7.62e$-02$ & 9.54e$-02$ \\
  3 & 27 & 8.73e$-03$ & 2.38e$-02$ & 6.74e$-02$ & 1.90e$-01$ & 3.34e$-01$ & 4.91e$-01$ & 6.57e$-01$ & 1.11e$+00$ & 1.62e$+00$ \\
  3 & 28 & 2.92e$-03$ & 7.27e$-03$ & 1.70e$-02$ & 3.57e$-02$ & 5.41e$-02$ & 7.47e$-02$ & 9.87e$-02$ & 1.75e$-01$ & 2.72e$-01$ \\
  3 & 29 & 9.04e$-05$ & 5.21e$-04$ & 3.04e$-03$ & 1.66e$-02$ & 4.17e$-02$ & 7.71e$-02$ & 1.21e$-01$ & 2.63e$-01$ & 4.40e$-01$ \\
  3 & 30 & 3.80e$-03$ & 1.21e$-02$ & 3.86e$-02$ & 1.12e$-01$ & 1.97e$-01$ & 2.82e$-01$ & 3.66e$-01$ & 5.62e$-01$ & 7.37e$-01$ \\
  3 & 31 & 3.27e$-03$ & 9.64e$-03$ & 2.84e$-02$ & 8.45e$-02$ & 1.58e$-01$ & 2.39e$-01$ & 3.23e$-01$ & 5.32e$-01$ & 7.25e$-01$ \\
  3 & 32 & 3.10e$-03$ & 9.18e$-03$ & 2.68e$-02$ & 6.95e$-02$ & 1.14e$-01$ & 1.56e$-01$ & 1.97e$-01$ & 2.90e$-01$ & 3.70e$-01$ \\
  3 & 33 & 2.91e$-03$ & 8.85e$-03$ & 3.23e$-02$ & 1.29e$-01$ & 2.88e$-01$ & 4.99e$-01$ & 7.53e$-01$ & 1.53e$+00$ & 2.44e$+00$ \\
  3 & 34 & 2.20e$-03$ & 6.47e$-03$ & 2.23e$-02$ & 8.90e$-02$ & 1.98e$-01$ & 3.38e$-01$ & 5.02e$-01$ & 9.82e$-01$ & 1.53e$+00$ \\
  3 & 35 & 8.28e$-04$ & 2.25e$-03$ & 6.94e$-03$ & 2.38e$-02$ & 4.57e$-02$ & 6.88e$-02$ & 9.18e$-02$ & 1.46e$-01$ & 1.95e$-01$ \\
  3 & 36 & 7.02e$-04$ & 2.30e$-03$ & 8.66e$-03$ & 3.21e$-02$ & 6.83e$-02$ & 1.17e$-01$ & 1.77e$-01$ & 3.64e$-01$ & 5.87e$-01$ \\
  3 & 37 & 4.85e$-04$ & 1.77e$-03$ & 7.92e$-03$ & 4.61e$-02$ & 1.38e$-01$ & 2.95e$-01$ & 5.20e$-01$ & 1.38e$+00$ & 2.62e$+00$ \\
  4 &  5 & 1.37e$-02$ & 2.85e$-02$ & 6.84e$-02$ & 1.89e$-01$ & 3.36e$-01$ & 4.92e$-01$ & 6.52e$-01$ & 1.06e$+00$ & 1.46e$+00$ \\
  4 &  6 & 1.12e$-01$ & 2.76e$-01$ & 6.33e$-01$ & 1.35e$+00$ & 1.99e$+00$ & 2.56e$+00$ & 3.07e$+00$ & 4.13e$+00$ & 5.00e$+00$ \\
  4 &  9 & 7.29e$-03$ & 2.34e$-02$ & 7.23e$-02$ & 1.92e$-01$ & 3.14e$-01$ & 4.31e$-01$ & 5.45e$-01$ & 8.19e$-01$ & 1.08e$+00$ \\
  4 & 11 & 3.75e$-03$ & 1.03e$-02$ & 2.83e$-02$ & 6.63e$-02$ & 1.00e$-01$ & 1.30e$-01$ & 1.56e$-01$ & 2.12e$-01$ & 2.60e$-01$ \\
  4 & 12 & 1.84e$-02$ & 4.22e$-02$ & 9.38e$-02$ & 2.02e$-01$ & 3.11e$-01$ & 4.21e$-01$ & 5.30e$-01$ & 7.89e$-01$ & 1.03e$+00$ \\
  4 & 15 & 1.10e$-02$ & 2.48e$-02$ & 5.71e$-02$ & 1.36e$-01$ & 2.24e$-01$ & 3.15e$-01$ & 4.06e$-01$ & 6.30e$-01$ & 8.36e$-01$ \\
  4 & 17 & 3.39e$-03$ & 1.02e$-02$ & 2.62e$-02$ & 5.61e$-02$ & 8.37e$-02$ & 1.11e$-01$ & 1.37e$-01$ & 2.05e$-01$ & 2.71e$-01$ \\
  4 & 19 & 3.46e$-03$ & 8.55e$-03$ & 1.81e$-02$ & 3.33e$-02$ & 4.50e$-02$ & 5.48e$-02$ & 6.34e$-02$ & 8.23e$-02$ & 9.90e$-02$ \\
  4 & 20 & 6.93e$-03$ & 1.77e$-02$ & 4.10e$-02$ & 9.30e$-02$ & 1.53e$-01$ & 2.18e$-01$ & 2.84e$-01$ & 4.40e$-01$ & 5.81e$-01$ \\
  4 & 23 & 5.85e$-03$ & 1.35e$-02$ & 3.07e$-02$ & 7.42e$-02$ & 1.28e$-01$ & 1.88e$-01$ & 2.50e$-01$ & 3.99e$-01$ & 5.34e$-01$ \\
  4 & 25 & 2.90e$-03$ & 7.92e$-03$ & 1.88e$-02$ & 4.00e$-02$ & 5.96e$-02$ & 7.79e$-02$ & 9.52e$-02$ & 1.35e$-01$ & 1.72e$-01$ \\
  4 & 27 & 5.25e$-03$ & 1.49e$-02$ & 4.33e$-02$ & 1.17e$-01$ & 1.93e$-01$ & 2.64e$-01$ & 3.29e$-01$ & 4.64e$-01$ & 5.72e$-01$ \\
  4 & 28 & 2.43e$-03$ & 5.50e$-03$ & 1.13e$-02$ & 2.06e$-02$ & 2.76e$-02$ & 3.34e$-02$ & 3.85e$-02$ & 5.00e$-02$ & 6.06e$-02$ \\
  4 & 33 & 2.86e$-03$ & 8.27e$-03$ & 2.65e$-02$ & 8.32e$-02$ & 1.50e$-01$ & 2.19e$-01$ & 2.85e$-01$ & 4.35e$-01$ & 5.61e$-01$ \\
  4 & 34 & 1.58e$-03$ & 4.69e$-03$ & 1.37e$-02$ & 3.53e$-02$ & 5.73e$-02$ & 7.80e$-02$ & 9.74e$-02$ & 1.41e$-01$ & 1.81e$-01$ \\
  4 & 36 & 6.88e$-04$ & 1.71e$-03$ & 4.44e$-03$ & 1.19e$-02$ & 2.06e$-02$ & 3.01e$-02$ & 3.98e$-02$ & 6.52e$-02$ & 9.06e$-02$ \\
  4 & 37 & 7.14e$-05$ & 4.19e$-04$ & 1.94e$-03$ & 6.23e$-03$ & 1.09e$-02$ & 1.55e$-02$ & 2.01e$-02$ & 3.10e$-02$ & 4.09e$-02$ \\
  5 &  6 & 2.91e$-02$ & 6.62e$-02$ & 1.64e$-01$ & 4.61e$-01$ & 8.76e$-01$ & 1.39e$+00$ & 1.98e$+00$ & 3.82e$+00$ & 6.17e$+00$ \\
  5 &  7 & 1.15e$-01$ & 2.48e$-01$ & 4.82e$-01$ & 8.39e$-01$ & 1.11e$+00$ & 1.32e$+00$ & 1.51e$+00$ & 1.92e$+00$ & 2.27e$+00$ \\
  5 &  8 & 2.95e$-02$ & 7.14e$-02$ & 1.93e$-01$ & 5.09e$-01$ & 8.35e$-01$ & 1.14e$+00$ & 1.43e$+00$ & 2.06e$+00$ & 2.62e$+00$ \\
  5 &  9 & 1.58e$-02$ & 6.98e$-02$ & 3.62e$-01$ & 1.63e$+00$ & 3.75e$+00$ & 6.71e$+00$ & 1.05e$+01$ & 2.38e$+01$ & 4.26e$+01$ \\
  5 & 10 & 2.91e$-03$ & 1.03e$-02$ & 3.10e$-02$ & 7.41e$-02$ & 1.13e$-01$ & 1.46e$-01$ & 1.76e$-01$ & 2.43e$-01$ & 3.01e$-01$ \\
  5 & 11 & 1.69e$-02$ & 4.30e$-02$ & 1.01e$-01$ & 2.21e$-01$ & 3.44e$-01$ & 4.79e$-01$ & 6.34e$-01$ & 1.14e$+00$ & 1.87e$+00$ \\
  5 & 12 & 2.72e$-02$ & 8.21e$-02$ & 2.35e$-01$ & 6.55e$-01$ & 1.21e$+00$ & 1.91e$+00$ & 2.78e$+00$ & 5.71e$+00$ & 9.64e$+00$ \\
  5 & 13 & 2.73e$-02$ & 5.83e$-02$ & 1.24e$-01$ & 2.54e$-01$ & 3.71e$-01$ & 4.76e$-01$ & 5.71e$-01$ & 7.74e$-01$ & 9.42e$-01$ \\
  5 & 14 & 1.38e$-02$ & 3.55e$-02$ & 7.92e$-02$ & 1.50e$-01$ & 2.06e$-01$ & 2.54e$-01$ & 2.97e$-01$ & 3.94e$-01$ & 4.81e$-01$ \\
  5 & 15 & 1.62e$-02$ & 4.69e$-02$ & 1.34e$-01$ & 4.39e$-01$ & 9.63e$-01$ & 1.71e$+00$ & 2.64e$+00$ & 5.64e$+00$ & 9.30e$+00$ \\
  5 & 16 & 8.17e$-03$ & 2.13e$-02$ & 6.02e$-02$ & 1.70e$-01$ & 2.96e$-01$ & 4.28e$-01$ & 5.58e$-01$ & 8.68e$-01$ & 1.14e$+00$ \\
  5 & 17 & 7.88e$-03$ & 2.82e$-02$ & 8.06e$-02$ & 2.02e$-01$ & 3.43e$-01$ & 5.13e$-01$ & 7.17e$-01$ & 1.40e$+00$ & 2.35e$+00$ \\
  5 & 18 & 1.93e$-03$ & 6.24e$-03$ & 1.47e$-02$ & 2.89e$-02$ & 4.10e$-02$ & 5.17e$-02$ & 6.15e$-02$ & 8.38e$-02$ & 1.04e$-01$ \\
  5 & 19 & 6.65e$-03$ & 1.57e$-02$ & 3.55e$-02$ & 7.82e$-02$ & 1.26e$-01$ & 1.80e$-01$ & 2.45e$-01$ & 4.57e$-01$ & 7.43e$-01$ \\
  5 & 20 & 9.56e$-03$ & 2.76e$-02$ & 6.66e$-02$ & 1.52e$-01$ & 2.62e$-01$ & 4.03e$-01$ & 5.75e$-01$ & 1.14e$+00$ & 1.89e$+00$ \\
  5 & 21 & 8.22e$-03$ & 2.05e$-02$ & 4.72e$-02$ & 1.06e$-01$ & 1.67e$-01$ & 2.28e$-01$ & 2.86e$-01$ & 4.17e$-01$ & 5.27e$-01$ \\
  5 & 22 & 6.70e$-03$ & 1.67e$-02$ & 3.53e$-02$ & 6.46e$-02$ & 8.95e$-02$ & 1.13e$-01$ & 1.37e$-01$ & 1.96e$-01$ & 2.52e$-01$ \\
  5 & 23 & 8.23e$-03$ & 2.11e$-02$ & 5.44e$-02$ & 1.58e$-01$ & 3.11e$-01$ & 5.08e$-01$ & 7.40e$-01$ & 1.44e$+00$ & 2.25e$+00$ \\
  5 & 24 & 3.32e$-03$ & 9.98e$-03$ & 2.72e$-02$ & 7.78e$-02$ & 1.45e$-01$ & 2.18e$-01$ & 2.92e$-01$ & 4.68e$-01$ & 6.22e$-01$ \\
  5 & 25 & 7.17e$-03$ & 1.90e$-02$ & 4.21e$-02$ & 8.53e$-02$ & 1.38e$-01$ & 2.09e$-01$ & 3.02e$-01$ & 6.27e$-01$ & 1.06e$+00$ \\
  5 & 26 & 2.63e$-03$ & 7.31e$-03$ & 1.67e$-02$ & 3.19e$-02$ & 4.30e$-02$ & 5.16e$-02$ & 5.88e$-02$ & 7.33e$-02$ & 8.50e$-02$ \\
  5 & 27 & 5.61e$-03$ & 1.66e$-02$ & 4.97e$-02$ & 1.42e$-01$ & 2.52e$-01$ & 3.69e$-01$ & 4.93e$-01$ & 8.20e$-01$ & 1.17e$+00$ \\
  5 & 28 & 3.82e$-03$ & 1.03e$-02$ & 2.66e$-02$ & 6.14e$-02$ & 9.67e$-02$ & 1.36e$-01$ & 1.80e$-01$ & 3.17e$-01$ & 4.93e$-01$ \\
  5 & 29 & 6.96e$-06$ & 4.24e$-05$ & 2.61e$-04$ & 1.65e$-03$ & 4.86e$-03$ & 1.03e$-02$ & 1.80e$-02$ & 4.59e$-02$ & 8.32e$-02$ \\
  5 & 30 & 4.40e$-03$ & 1.37e$-02$ & 4.20e$-02$ & 1.14e$-01$ & 1.86e$-01$ & 2.53e$-01$ & 3.13e$-01$ & 4.39e$-01$ & 5.38e$-01$ \\
  5 & 31 & 3.19e$-03$ & 8.75e$-03$ & 2.32e$-02$ & 5.74e$-02$ & 9.32e$-02$ & 1.29e$-01$ & 1.63e$-01$ & 2.41e$-01$ & 3.10e$-01$ \\
  5 & 32 & 3.61e$-03$ & 1.06e$-02$ & 3.22e$-02$ & 9.49e$-02$ & 1.64e$-01$ & 2.30e$-01$ & 2.92e$-01$ & 4.23e$-01$ & 5.29e$-01$ \\
  5 & 33 & 3.59e$-03$ & 1.01e$-02$ & 2.96e$-02$ & 9.21e$-02$ & 1.80e$-01$ & 2.87e$-01$ & 4.07e$-01$ & 7.50e$-01$ & 1.13e$+00$ \\
  5 & 34 & 2.20e$-03$ & 6.96e$-03$ & 2.80e$-02$ & 1.17e$-01$ & 2.59e$-01$ & 4.38e$-01$ & 6.45e$-01$ & 1.24e$+00$ & 1.89e$+00$ \\
  5 & 35 & 8.19e$-04$ & 2.16e$-03$ & 5.00e$-03$ & 1.17e$-02$ & 1.97e$-02$ & 2.83e$-02$ & 3.71e$-02$ & 5.87e$-02$ & 7.89e$-02$ \\
  5 & 36 & 1.00e$-03$ & 2.97e$-03$ & 1.01e$-02$ & 3.61e$-02$ & 7.75e$-02$ & 1.33e$-01$ & 2.01e$-01$ & 4.12e$-01$ & 6.65e$-01$ \\
  5 & 37 & 5.14e$-04$ & 1.79e$-03$ & 5.48e$-03$ & 1.69e$-02$ & 3.52e$-02$ & 6.05e$-02$ & 9.24e$-02$ & 1.94e$-01$ & 3.17e$-01$ \\
  6 &  7 & 6.76e$-02$ & 1.77e$-01$ & 4.98e$-01$ & 1.38e$+00$ & 2.35e$+00$ & 3.30e$+00$ & 4.19e$+00$ & 6.21e$+00$ & 7.96e$+00$ \\
  6 &  8 & 3.52e$-02$ & 1.19e$-01$ & 4.14e$-01$ & 1.33e$+00$ & 2.46e$+00$ & 3.69e$+00$ & 4.98e$+00$ & 8.34e$+00$ & 1.17e$+01$ \\
  6 &  9 & 6.19e$-02$ & 2.97e$-01$ & 1.29e$+00$ & 4.92e$+00$ & 1.04e$+01$ & 1.77e$+01$ & 2.65e$+01$ & 5.48e$+01$ & 9.13e$+01$ \\
  6 & 10 & 4.77e$-03$ & 1.82e$-02$ & 6.18e$-02$ & 1.73e$-01$ & 2.90e$-01$ & 4.06e$-01$ & 5.18e$-01$ & 7.80e$-01$ & 1.01e$+00$ \\
  6 & 11 & 9.99e$-02$ & 2.60e$-01$ & 6.27e$-01$ & 1.46e$+00$ & 2.38e$+00$ & 3.36e$+00$ & 4.40e$+00$ & 7.19e$+00$ & 1.02e$+01$ \\
  6 & 12 & 1.02e$-01$ & 3.09e$-01$ & 8.65e$-01$ & 2.36e$+00$ & 4.36e$+00$ & 6.88e$+00$ & 9.94e$+00$ & 1.98e$+01$ & 3.24e$+01$ \\
  6 & 13 & 5.16e$-02$ & 1.22e$-01$ & 2.76e$-01$ & 6.01e$-01$ & 9.25e$-01$ & 1.24e$+00$ & 1.55e$+00$ & 2.28e$+00$ & 2.95e$+00$ \\
  6 & 14 & 2.33e$-02$ & 5.83e$-02$ & 1.49e$-01$ & 3.50e$-01$ & 5.41e$-01$ & 7.17e$-01$ & 8.80e$-01$ & 1.25e$+00$ & 1.56e$+00$ \\
  6 & 15 & 5.71e$-02$ & 1.62e$-01$ & 4.60e$-01$ & 1.45e$+00$ & 3.08e$+00$ & 5.45e$+00$ & 8.61e$+00$ & 2.01e$+01$ & 3.65e$+01$ \\
  6 & 16 & 2.01e$-02$ & 5.05e$-02$ & 1.40e$-01$ & 3.90e$-01$ & 6.96e$-01$ & 1.04e$+00$ & 1.40e$+00$ & 2.32e$+00$ & 3.21e$+00$ \\
  6 & 17 & 2.32e$-02$ & 8.25e$-02$ & 2.49e$-01$ & 6.79e$-01$ & 1.20e$+00$ & 1.80e$+00$ & 2.47e$+00$ & 4.43e$+00$ & 6.75e$+00$ \\
  6 & 18 & 5.67e$-03$ & 1.80e$-02$ & 4.50e$-02$ & 9.69e$-02$ & 1.46e$-01$ & 1.93e$-01$ & 2.38e$-01$ & 3.41e$-01$ & 4.30e$-01$ \\
  6 & 19 & 2.15e$-02$ & 5.30e$-02$ & 1.18e$-01$ & 2.58e$-01$ & 4.20e$-01$ & 6.06e$-01$ & 8.14e$-01$ & 1.41e$+00$ & 2.06e$+00$ \\
  6 & 20 & 4.29e$-02$ & 1.18e$-01$ & 3.01e$-01$ & 8.36e$-01$ & 1.62e$+00$ & 2.64e$+00$ & 3.84e$+00$ & 7.47e$+00$ & 1.17e$+01$ \\
  6 & 21 & 2.01e$-02$ & 5.16e$-02$ & 1.17e$-01$ & 2.63e$-01$ & 4.34e$-01$ & 6.17e$-01$ & 8.03e$-01$ & 1.26e$+00$ & 1.68e$+00$ \\
  6 & 22 & 1.06e$-02$ & 2.76e$-02$ & 6.46e$-02$ & 1.43e$-01$ & 2.30e$-01$ & 3.23e$-01$ & 4.19e$-01$ & 6.58e$-01$ & 8.81e$-01$ \\
  6 & 23 & 2.66e$-02$ & 7.30e$-02$ & 1.88e$-01$ & 5.38e$-01$ & 1.10e$+00$ & 1.90e$+00$ & 2.94e$+00$ & 6.60e$+00$ & 1.16e$+01$ \\
  6 & 24 & 1.10e$-02$ & 3.09e$-02$ & 8.10e$-02$ & 2.10e$-01$ & 3.71e$-01$ & 5.49e$-01$ & 7.33e$-01$ & 1.19e$+00$ & 1.61e$+00$ \\
  6 & 25 & 1.38e$-02$ & 4.34e$-02$ & 1.19e$-01$ & 2.97e$-01$ & 5.09e$-01$ & 7.55e$-01$ & 1.03e$+00$ & 1.82e$+00$ & 2.73e$+00$ \\
  6 & 26 & 5.83e$-03$ & 1.67e$-02$ & 4.16e$-02$ & 8.85e$-02$ & 1.27e$-01$ & 1.61e$-01$ & 1.90e$-01$ & 2.49e$-01$ & 2.97e$-01$ \\
  6 & 27 & 2.05e$-02$ & 6.53e$-02$ & 2.21e$-01$ & 7.26e$-01$ & 1.38e$+00$ & 2.12e$+00$ & 2.89e$+00$ & 4.92e$+00$ & 6.97e$+00$ \\
  6 & 28 & 8.81e$-03$ & 2.61e$-02$ & 6.84e$-02$ & 1.55e$-01$ & 2.45e$-01$ & 3.44e$-01$ & 4.54e$-01$ & 7.60e$-01$ & 1.09e$+00$ \\
  6 & 29 & 3.83e$-05$ & 2.30e$-04$ & 1.43e$-03$ & 8.48e$-03$ & 2.25e$-02$ & 4.29e$-02$ & 6.90e$-02$ & 1.54e$-01$ & 2.60e$-01$ \\
  6 & 30 & 1.20e$-02$ & 3.67e$-02$ & 1.11e$-01$ & 3.04e$-01$ & 5.12e$-01$ & 7.15e$-01$ & 9.07e$-01$ & 1.34e$+00$ & 1.70e$+00$ \\
  6 & 31 & 7.67e$-03$ & 2.20e$-02$ & 6.27e$-02$ & 1.75e$-01$ & 3.09e$-01$ & 4.49e$-01$ & 5.86e$-01$ & 9.04e$-01$ & 1.18e$+00$ \\
  6 & 32 & 9.56e$-03$ & 2.81e$-02$ & 8.63e$-02$ & 2.43e$-01$ & 4.08e$-01$ & 5.64e$-01$ & 7.10e$-01$ & 1.03e$+00$ & 1.30e$+00$ \\
  6 & 33 & 1.48e$-02$ & 4.11e$-02$ & 1.29e$-01$ & 4.72e$-01$ & 1.02e$+00$ & 1.76e$+00$ & 2.67e$+00$ & 5.54e$+00$ & 9.16e$+00$ \\
  6 & 34 & 5.76e$-03$ & 2.00e$-02$ & 7.70e$-02$ & 2.95e$-01$ & 6.12e$-01$ & 9.90e$-01$ & 1.41e$+00$ & 2.58e$+00$ & 3.85e$+00$ \\
  6 & 35 & 1.76e$-03$ & 4.98e$-03$ & 1.46e$-02$ & 4.89e$-02$ & 9.65e$-02$ & 1.48e$-01$ & 1.99e$-01$ & 3.16e$-01$ & 4.16e$-01$ \\
  6 & 36 & 1.99e$-03$ & 5.91e$-03$ & 2.06e$-02$ & 7.83e$-02$ & 1.66e$-01$ & 2.78e$-01$ & 4.14e$-01$ & 8.42e$-01$ & 1.36e$+00$ \\
  6 & 37 & 7.19e$-04$ & 2.80e$-03$ & 1.01e$-02$ & 3.57e$-02$ & 7.64e$-02$ & 1.31e$-01$ & 1.98e$-01$ & 4.07e$-01$ & 6.59e$-01$ \\
  7 &  9 & 2.04e$-02$ & 6.32e$-02$ & 1.97e$-01$ & 5.71e$-01$ & 1.01e$+00$ & 1.50e$+00$ & 2.01e$+00$ & 3.35e$+00$ & 4.69e$+00$ \\
  7 & 11 & 1.48e$-02$ & 3.51e$-02$ & 7.96e$-02$ & 1.69e$-01$ & 2.58e$-01$ & 3.46e$-01$ & 4.33e$-01$ & 6.38e$-01$ & 8.24e$-01$ \\
  7 & 12 & 7.97e$-02$ & 1.90e$-01$ & 4.24e$-01$ & 8.96e$-01$ & 1.35e$+00$ & 1.78e$+00$ & 2.20e$+00$ & 3.14e$+00$ & 3.97e$+00$ \\
  7 & 15 & 3.01e$-02$ & 8.15e$-02$ & 2.09e$-01$ & 5.24e$-01$ & 8.92e$-01$ & 1.29e$+00$ & 1.70e$+00$ & 2.73e$+00$ & 3.70e$+00$ \\
  7 & 17 & 1.14e$-02$ & 3.29e$-02$ & 8.76e$-02$ & 2.18e$-01$ & 3.66e$-01$ & 5.18e$-01$ & 6.68e$-01$ & 1.02e$+00$ & 1.32e$+00$ \\
  7 & 19 & 9.17e$-03$ & 2.36e$-02$ & 5.39e$-02$ & 1.09e$-01$ & 1.56e$-01$ & 1.98e$-01$ & 2.36e$-01$ & 3.21e$-01$ & 3.95e$-01$ \\
  7 & 20 & 2.04e$-02$ & 5.49e$-02$ & 1.32e$-01$ & 3.01e$-01$ & 4.88e$-01$ & 6.85e$-01$ & 8.81e$-01$ & 1.34e$+00$ & 1.76e$+00$ \\
  7 & 23 & 1.45e$-02$ & 3.83e$-02$ & 8.85e$-02$ & 1.98e$-01$ & 3.22e$-01$ & 4.58e$-01$ & 6.00e$-01$ & 9.51e$-01$ & 1.28e$+00$ \\
  7 & 25 & 7.82e$-03$ & 2.42e$-02$ & 6.61e$-02$ & 1.59e$-01$ & 2.51e$-01$ & 3.37e$-01$ & 4.17e$-01$ & 5.93e$-01$ & 7.39e$-01$ \\
  7 & 27 & 1.47e$-02$ & 4.13e$-02$ & 1.21e$-01$ & 3.32e$-01$ & 5.52e$-01$ & 7.57e$-01$ & 9.41e$-01$ & 1.33e$+00$ & 1.63e$+00$ \\
  7 & 28 & 5.49e$-03$ & 1.44e$-02$ & 3.48e$-02$ & 7.32e$-02$ & 1.06e$-01$ & 1.34e$-01$ & 1.60e$-01$ & 2.14e$-01$ & 2.59e$-01$ \\
  7 & 33 & 9.10e$-03$ & 2.40e$-02$ & 6.24e$-02$ & 1.66e$-01$ & 2.88e$-01$ & 4.17e$-01$ & 5.44e$-01$ & 8.47e$-01$ & 1.12e$+00$ \\
  7 & 34 & 4.25e$-03$ & 1.36e$-02$ & 4.10e$-02$ & 1.15e$-01$ & 2.00e$-01$ & 2.87e$-01$ & 3.72e$-01$ & 5.68e$-01$ & 7.37e$-01$ \\
  7 & 36 & 1.96e$-03$ & 5.04e$-03$ & 1.26e$-02$ & 3.25e$-02$ & 5.69e$-02$ & 8.35e$-02$ & 1.11e$-01$ & 1.79e$-01$ & 2.43e$-01$ \\
  7 & 37 & 6.48e$-04$ & 2.70e$-03$ & 9.59e$-03$ & 2.92e$-02$ & 5.17e$-02$ & 7.46e$-02$ & 9.68e$-02$ & 1.48e$-01$ & 1.93e$-01$ \\
  8 &  9 & 2.53e$-02$ & 8.86e$-02$ & 3.13e$-01$ & 8.95e$-01$ & 1.47e$+00$ & 2.00e$+00$ & 2.50e$+00$ & 3.63e$+00$ & 4.65e$+00$ \\
  8 & 11 & 3.33e$-02$ & 8.51e$-02$ & 1.85e$-01$ & 3.49e$-01$ & 4.83e$-01$ & 5.99e$-01$ & 7.03e$-01$ & 9.29e$-01$ & 1.12e$+00$ \\
  8 & 12 & 1.41e$-01$ & 3.41e$-01$ & 7.29e$-01$ & 1.40e$+00$ & 1.99e$+00$ & 2.54e$+00$ & 3.07e$+00$ & 4.29e$+00$ & 5.38e$+00$ \\
  8 & 15 & 5.93e$-02$ & 1.49e$-01$ & 3.81e$-01$ & 9.28e$-01$ & 1.51e$+00$ & 2.11e$+00$ & 2.71e$+00$ & 4.17e$+00$ & 5.53e$+00$ \\
  8 & 17 & 2.35e$-02$ & 7.25e$-02$ & 1.97e$-01$ & 4.55e$-01$ & 6.93e$-01$ & 9.09e$-01$ & 1.11e$+00$ & 1.54e$+00$ & 1.92e$+00$ \\
  8 & 19 & 1.32e$-02$ & 3.31e$-02$ & 7.45e$-02$ & 1.48e$-01$ & 2.08e$-01$ & 2.60e$-01$ & 3.05e$-01$ & 4.02e$-01$ & 4.81e$-01$ \\
  8 & 20 & 3.22e$-02$ & 8.54e$-02$ & 1.97e$-01$ & 4.26e$-01$ & 6.71e$-01$ & 9.26e$-01$ & 1.18e$+00$ & 1.79e$+00$ & 2.34e$+00$ \\
  8 & 23 & 2.78e$-02$ & 7.33e$-02$ & 1.76e$-01$ & 4.12e$-01$ & 6.77e$-01$ & 9.54e$-01$ & 1.23e$+00$ & 1.90e$+00$ & 2.51e$+00$ \\
  8 & 25 & 1.75e$-02$ & 5.06e$-02$ & 1.25e$-01$ & 2.69e$-01$ & 3.97e$-01$ & 5.13e$-01$ & 6.18e$-01$ & 8.49e$-01$ & 1.05e$+00$ \\
  8 & 27 & 2.50e$-02$ & 7.39e$-02$ & 2.04e$-01$ & 5.07e$-01$ & 8.04e$-01$ & 1.08e$+00$ & 1.33e$+00$ & 1.86e$+00$ & 2.30e$+00$ \\
  8 & 28 & 7.32e$-03$ & 1.93e$-02$ & 4.81e$-02$ & 1.03e$-01$ & 1.48e$-01$ & 1.87e$-01$ & 2.20e$-01$ & 2.89e$-01$ & 3.44e$-01$ \\
  8 & 33 & 1.34e$-02$ & 4.36e$-02$ & 1.37e$-01$ & 3.87e$-01$ & 6.55e$-01$ & 9.18e$-01$ & 1.17e$+00$ & 1.74e$+00$ & 2.23e$+00$ \\
  8 & 34 & 6.57e$-03$ & 2.24e$-02$ & 6.98e$-02$ & 2.03e$-01$ & 3.55e$-01$ & 5.06e$-01$ & 6.50e$-01$ & 9.72e$-01$ & 1.25e$+00$ \\
  8 & 36 & 3.10e$-03$ & 8.81e$-03$ & 2.50e$-02$ & 7.54e$-02$ & 1.41e$-01$ & 2.14e$-01$ & 2.89e$-01$ & 4.67e$-01$ & 6.24e$-01$ \\
  8 & 37 & 9.05e$-04$ & 4.23e$-03$ & 1.61e$-02$ & 5.15e$-02$ & 9.49e$-02$ & 1.41e$-01$ & 1.87e$-01$ & 2.94e$-01$ & 3.88e$-01$ \\
  9 & 10 & 3.72e$-02$ & 9.04e$-02$ & 2.05e$-01$ & 3.98e$-01$ & 5.48e$-01$ & 6.76e$-01$ & 7.90e$-01$ & 1.04e$+00$ & 1.25e$+00$ \\
  9 & 11 & 1.28e$-01$ & 3.76e$-01$ & 1.15e$+00$ & 3.79e$+00$ & 7.97e$+00$ & 1.37e$+01$ & 2.11e$+01$ & 4.64e$+01$ & 8.16e$+01$ \\
  9 & 12 & 2.29e$-01$ & 6.34e$-01$ & 1.67e$+00$ & 4.52e$+00$ & 8.78e$+00$ & 1.51e$+01$ & 2.38e$+01$ & 5.82e$+01$ & 1.12e$+02$ \\
  9 & 13 & 1.66e$-01$ & 3.62e$-01$ & 7.36e$-01$ & 1.39e$+00$ & 1.95e$+00$ & 2.43e$+00$ & 2.85e$+00$ & 3.73e$+00$ & 4.44e$+00$ \\
  9 & 14 & 7.73e$-02$ & 1.91e$-01$ & 4.50e$-01$ & 9.54e$-01$ & 1.39e$+00$ & 1.77e$+00$ & 2.10e$+00$ & 2.78e$+00$ & 3.32e$+00$ \\
  9 & 15 & 1.12e$-01$ & 3.44e$-01$ & 1.05e$+00$ & 3.68e$+00$ & 8.16e$+00$ & 1.44e$+01$ & 2.23e$+01$ & 4.80e$+01$ & 8.03e$+01$ \\
  9 & 16 & 5.39e$-02$ & 1.36e$-01$ & 3.51e$-01$ & 8.65e$-01$ & 1.42e$+00$ & 1.98e$+00$ & 2.53e$+00$ & 3.80e$+00$ & 4.93e$+00$ \\
  9 & 17 & 7.63e$-02$ & 2.56e$-01$ & 7.88e$-01$ & 2.30e$+00$ & 4.33e$+00$ & 6.86e$+00$ & 9.85e$+00$ & 1.92e$+01$ & 3.11e$+01$ \\
  9 & 18 & 1.10e$-02$ & 3.43e$-02$ & 8.36e$-02$ & 1.70e$-01$ & 2.41e$-01$ & 3.02e$-01$ & 3.56e$-01$ & 4.70e$-01$ & 5.61e$-01$ \\
  9 & 19 & 2.75e$-02$ & 7.55e$-02$ & 1.81e$-01$ & 4.16e$-01$ & 6.92e$-01$ & 1.03e$+00$ & 1.44e$+00$ & 2.85e$+00$ & 4.83e$+00$ \\
  9 & 20 & 8.88e$-02$ & 2.36e$-01$ & 5.52e$-01$ & 1.23e$+00$ & 2.10e$+00$ & 3.22e$+00$ & 4.60e$+00$ & 9.08e$+00$ & 1.47e$+01$ \\
  9 & 21 & 4.20e$-02$ & 1.04e$-01$ & 2.26e$-01$ & 4.51e$-01$ & 6.63e$-01$ & 8.64e$-01$ & 1.05e$+00$ & 1.47e$+00$ & 1.82e$+00$ \\
  9 & 22 & 2.39e$-02$ & 6.18e$-02$ & 1.33e$-01$ & 2.53e$-01$ & 3.63e$-01$ & 4.66e$-01$ & 5.65e$-01$ & 7.92e$-01$ & 9.89e$-01$ \\
  9 & 23 & 5.36e$-02$ & 1.54e$-01$ & 4.24e$-01$ & 1.25e$+00$ & 2.37e$+00$ & 3.70e$+00$ & 5.17e$+00$ & 9.28e$+00$ & 1.37e$+01$ \\
  9 & 24 & 2.31e$-02$ & 6.48e$-02$ & 1.55e$-01$ & 3.45e$-01$ & 5.46e$-01$ & 7.52e$-01$ & 9.55e$-01$ & 1.43e$+00$ & 1.85e$+00$ \\
  9 & 25 & 3.65e$-02$ & 1.07e$-01$ & 2.74e$-01$ & 6.52e$-01$ & 1.14e$+00$ & 1.75e$+00$ & 2.51e$+00$ & 4.94e$+00$ & 7.98e$+00$ \\
  9 & 26 & 1.07e$-02$ & 3.01e$-02$ & 6.96e$-02$ & 1.36e$-01$ & 1.87e$-01$ & 2.26e$-01$ & 2.59e$-01$ & 3.23e$-01$ & 3.70e$-01$ \\
  9 & 27 & 3.43e$-02$ & 1.05e$-01$ & 3.20e$-01$ & 9.28e$-01$ & 1.69e$+00$ & 2.56e$+00$ & 3.51e$+00$ & 6.09e$+00$ & 8.85e$+00$ \\
  9 & 28 & 1.40e$-02$ & 4.14e$-02$ & 1.17e$-01$ & 2.92e$-01$ & 4.79e$-01$ & 6.89e$-01$ & 9.30e$-01$ & 1.69e$+00$ & 2.66e$+00$ \\
  9 & 29 & 5.96e$-06$ & 3.55e$-05$ & 2.11e$-04$ & 1.34e$-03$ & 3.91e$-03$ & 7.97e$-03$ & 1.34e$-02$ & 3.16e$-02$ & 5.51e$-02$ \\
  9 & 30 & 1.64e$-02$ & 5.12e$-02$ & 1.47e$-01$ & 3.67e$-01$ & 5.83e$-01$ & 7.84e$-01$ & 9.68e$-01$ & 1.36e$+00$ & 1.67e$+00$ \\
  9 & 31 & 1.19e$-02$ & 3.47e$-02$ & 9.80e$-02$ & 2.54e$-01$ & 4.16e$-01$ & 5.71e$-01$ & 7.13e$-01$ & 1.02e$+00$ & 1.27e$+00$ \\
  9 & 32 & 1.51e$-02$ & 4.58e$-02$ & 1.29e$-01$ & 3.26e$-01$ & 5.21e$-01$ & 7.01e$-01$ & 8.64e$-01$ & 1.21e$+00$ & 1.48e$+00$ \\
  9 & 33 & 2.10e$-02$ & 7.05e$-02$ & 2.45e$-01$ & 8.33e$-01$ & 1.64e$+00$ & 2.57e$+00$ & 3.58e$+00$ & 6.28e$+00$ & 9.10e$+00$ \\
  9 & 34 & 1.15e$-02$ & 4.03e$-02$ & 1.64e$-01$ & 6.82e$-01$ & 1.49e$+00$ & 2.50e$+00$ & 3.64e$+00$ & 6.86e$+00$ & 1.03e$+01$ \\
  9 & 35 & 3.47e$-03$ & 9.55e$-03$ & 2.40e$-02$ & 5.99e$-02$ & 9.91e$-02$ & 1.38e$-01$ & 1.74e$-01$ & 2.54e$-01$ & 3.22e$-01$ \\
  9 & 36 & 4.52e$-03$ & 1.33e$-02$ & 4.01e$-02$ & 1.37e$-01$ & 2.94e$-01$ & 5.05e$-01$ & 7.64e$-01$ & 1.58e$+00$ & 2.59e$+00$ \\
  9 & 37 & 5.22e$-04$ & 2.24e$-03$ & 8.77e$-03$ & 3.37e$-02$ & 7.48e$-02$ & 1.32e$-01$ & 2.05e$-01$ & 4.48e$-01$ & 7.65e$-01$ \\
 10 & 11 & 2.78e$-02$ & 7.51e$-02$ & 1.67e$-01$ & 3.35e$-01$ & 4.85e$-01$ & 6.24e$-01$ & 7.52e$-01$ & 1.04e$+00$ & 1.29e$+00$ \\
 10 & 12 & 1.38e$-01$ & 3.24e$-01$ & 6.56e$-01$ & 1.14e$+00$ & 1.50e$+00$ & 1.80e$+00$ & 2.06e$+00$ & 2.60e$+00$ & 3.02e$+00$ \\
 10 & 15 & 6.46e$-02$ & 1.66e$-01$ & 4.03e$-01$ & 8.81e$-01$ & 1.31e$+00$ & 1.69e$+00$ & 2.04e$+00$ & 2.74e$+00$ & 3.30e$+00$ \\
 10 & 17 & 1.90e$-02$ & 5.80e$-02$ & 1.52e$-01$ & 3.32e$-01$ & 4.91e$-01$ & 6.33e$-01$ & 7.62e$-01$ & 1.03e$+00$ & 1.25e$+00$ \\
 10 & 19 & 9.00e$-03$ & 2.35e$-02$ & 5.41e$-02$ & 1.09e$-01$ & 1.56e$-01$ & 1.96e$-01$ & 2.30e$-01$ & 3.00e$-01$ & 3.58e$-01$ \\
 10 & 20 & 2.93e$-02$ & 6.96e$-02$ & 1.49e$-01$ & 2.86e$-01$ & 4.04e$-01$ & 5.09e$-01$ & 6.04e$-01$ & 8.02e$-01$ & 9.59e$-01$ \\
 10 & 23 & 3.37e$-02$ & 7.88e$-02$ & 1.61e$-01$ & 3.01e$-01$ & 4.23e$-01$ & 5.33e$-01$ & 6.33e$-01$ & 8.45e$-01$ & 1.01e$+00$ \\
 10 & 25 & 1.19e$-02$ & 3.44e$-02$ & 8.50e$-02$ & 1.72e$-01$ & 2.40e$-01$ & 2.97e$-01$ & 3.47e$-01$ & 4.48e$-01$ & 5.28e$-01$ \\
 10 & 27 & 1.77e$-02$ & 4.66e$-02$ & 1.11e$-01$ & 2.30e$-01$ & 3.26e$-01$ & 4.05e$-01$ & 4.72e$-01$ & 6.04e$-01$ & 7.02e$-01$ \\
 10 & 28 & 4.56e$-03$ & 1.27e$-02$ & 2.84e$-02$ & 5.42e$-02$ & 7.45e$-02$ & 9.16e$-02$ & 1.07e$-01$ & 1.40e$-01$ & 1.70e$-01$ \\
 10 & 33 & 1.29e$-02$ & 3.67e$-02$ & 9.07e$-02$ & 2.01e$-01$ & 3.04e$-01$ & 3.96e$-01$ & 4.78e$-01$ & 6.48e$-01$ & 7.77e$-01$ \\
 10 & 34 & 6.35e$-03$ & 1.82e$-02$ & 4.54e$-02$ & 1.03e$-01$ & 1.60e$-01$ & 2.13e$-01$ & 2.63e$-01$ & 3.72e$-01$ & 4.62e$-01$ \\
 10 & 36 & 2.06e$-03$ & 5.48e$-03$ & 1.29e$-02$ & 2.94e$-02$ & 4.83e$-02$ & 6.97e$-02$ & 9.30e$-02$ & 1.56e$-01$ & 2.18e$-01$ \\
 10 & 37 & 5.40e$-04$ & 2.14e$-03$ & 7.17e$-03$ & 2.04e$-02$ & 3.38e$-02$ & 4.60e$-02$ & 5.67e$-02$ & 7.78e$-02$ & 9.34e$-02$ \\
 11 & 12 & 1.66e$-01$ & 4.86e$-01$ & 1.29e$+00$ & 3.23e$+00$ & 5.60e$+00$ & 8.53e$+00$ & 1.21e$+01$ & 2.43e$+01$ & 4.12e$+01$ \\
 11 & 13 & 9.06e$-02$ & 2.01e$-01$ & 4.32e$-01$ & 8.54e$-01$ & 1.22e$+00$ & 1.53e$+00$ & 1.81e$+00$ & 2.40e$+00$ & 2.85e$+00$ \\
 11 & 14 & 5.07e$-02$ & 1.14e$-01$ & 2.26e$-01$ & 4.16e$-01$ & 5.86e$-01$ & 7.43e$-01$ & 8.88e$-01$ & 1.20e$+00$ & 1.46e$+00$ \\
 11 & 15 & 6.71e$-02$ & 1.98e$-01$ & 6.38e$-01$ & 2.04e$+00$ & 3.97e$+00$ & 6.32e$+00$ & 9.04e$+00$ & 1.72e$+01$ & 2.69e$+01$ \\
 11 & 16 & 2.42e$-02$ & 9.04e$-02$ & 2.86e$-01$ & 6.96e$-01$ & 1.06e$+00$ & 1.37e$+00$ & 1.65e$+00$ & 2.23e$+00$ & 2.68e$+00$ \\
 11 & 17 & 8.66e$-02$ & 3.69e$-01$ & 1.49e$+00$ & 6.63e$+00$ & 1.63e$+01$ & 3.08e$+01$ & 5.02e$+01$ & 1.21e$+02$ & 2.23e$+02$ \\
 11 & 18 & 7.39e$-03$ & 2.13e$-02$ & 5.09e$-02$ & 1.01e$-01$ & 1.42e$-01$ & 1.78e$-01$ & 2.09e$-01$ & 2.77e$-01$ & 3.37e$-01$ \\
 11 & 19 & 3.18e$-02$ & 9.19e$-02$ & 2.09e$-01$ & 4.38e$-01$ & 6.93e$-01$ & 1.01e$+00$ & 1.42e$+00$ & 3.01e$+00$ & 5.54e$+00$ \\
 11 & 20 & 1.14e$-01$ & 3.60e$-01$ & 9.79e$-01$ & 2.50e$+00$ & 4.67e$+00$ & 7.78e$+00$ & 1.19e$+01$ & 2.62e$+01$ & 4.55e$+01$ \\
 11 & 21 & 3.63e$-02$ & 9.53e$-02$ & 2.03e$-01$ & 3.65e$-01$ & 4.91e$-01$ & 5.99e$-01$ & 6.94e$-01$ & 8.90e$-01$ & 1.04e$+00$ \\
 11 & 22 & 2.54e$-02$ & 5.98e$-02$ & 1.16e$-01$ & 1.97e$-01$ & 2.63e$-01$ & 3.21e$-01$ & 3.74e$-01$ & 4.89e$-01$ & 5.87e$-01$ \\
 11 & 23 & 5.62e$-02$ & 1.89e$-01$ & 6.29e$-01$ & 2.41e$+00$ & 5.31e$+00$ & 9.01e$+00$ & 1.33e$+01$ & 2.54e$+01$ & 3.84e$+01$ \\
 11 & 24 & 2.55e$-02$ & 6.97e$-02$ & 1.65e$-01$ & 3.35e$-01$ & 4.79e$-01$ & 6.04e$-01$ & 7.14e$-01$ & 9.38e$-01$ & 1.11e$+00$ \\
 11 & 25 & 3.16e$-02$ & 9.94e$-02$ & 2.73e$-01$ & 6.74e$-01$ & 1.19e$+00$ & 1.89e$+00$ & 2.81e$+00$ & 6.15e$+00$ & 1.09e$+01$ \\
 11 & 26 & 4.10e$-03$ & 1.28e$-02$ & 3.24e$-02$ & 6.49e$-02$ & 8.92e$-02$ & 1.08e$-01$ & 1.25e$-01$ & 1.58e$-01$ & 1.86e$-01$ \\
 11 & 27 & 3.15e$-02$ & 1.03e$-01$ & 2.81e$-01$ & 7.49e$-01$ & 1.40e$+00$ & 2.23e$+00$ & 3.24e$+00$ & 6.33e$+00$ & 1.01e$+01$ \\
 11 & 28 & 1.69e$-02$ & 5.10e$-02$ & 1.57e$-01$ & 4.28e$-01$ & 7.26e$-01$ & 1.05e$+00$ & 1.43e$+00$ & 2.58e$+00$ & 4.05e$+00$ \\
 11 & 29 & 3.45e$-06$ & 2.20e$-05$ & 1.43e$-04$ & 1.05e$-03$ & 3.33e$-03$ & 7.17e$-03$ & 1.24e$-02$ & 3.07e$-02$ & 5.41e$-02$ \\
 11 & 30 & 1.19e$-02$ & 3.42e$-02$ & 9.34e$-02$ & 2.17e$-01$ & 3.28e$-01$ & 4.25e$-01$ & 5.10e$-01$ & 6.81e$-01$ & 8.09e$-01$ \\
 11 & 31 & 7.28e$-03$ & 1.97e$-02$ & 4.89e$-02$ & 1.14e$-01$ & 1.80e$-01$ & 2.44e$-01$ & 3.05e$-01$ & 4.39e$-01$ & 5.50e$-01$ \\
 11 & 32 & 9.88e$-03$ & 3.02e$-02$ & 8.29e$-02$ & 1.98e$-01$ & 3.02e$-01$ & 3.92e$-01$ & 4.70e$-01$ & 6.24e$-01$ & 7.37e$-01$ \\
 11 & 33 & 2.71e$-02$ & 1.09e$-01$ & 3.92e$-01$ & 1.28e$+00$ & 2.44e$+00$ & 3.77e$+00$ & 5.19e$+00$ & 8.98e$+00$ & 1.29e$+01$ \\
 11 & 34 & 1.15e$-02$ & 5.23e$-02$ & 2.21e$-01$ & 8.73e$-01$ & 1.90e$+00$ & 3.24e$+00$ & 4.80e$+00$ & 9.41e$+00$ & 1.46e$+01$ \\
 11 & 35 & 1.88e$-03$ & 5.00e$-03$ & 1.19e$-02$ & 2.73e$-02$ & 4.53e$-02$ & 6.51e$-02$ & 8.59e$-02$ & 1.39e$-01$ & 1.91e$-01$ \\
 11 & 36 & 3.83e$-03$ & 1.32e$-02$ & 4.34e$-02$ & 1.39e$-01$ & 2.70e$-01$ & 4.30e$-01$ & 6.17e$-01$ & 1.20e$+00$ & 1.93e$+00$ \\
 11 & 37 & 6.06e$-04$ & 2.10e$-03$ & 6.28e$-03$ & 1.72e$-02$ & 3.28e$-02$ & 5.41e$-02$ & 8.10e$-02$ & 1.69e$-01$ & 2.78e$-01$ \\
 12 & 13 & 3.03e$-01$ & 7.41e$-01$ & 1.74e$+00$ & 3.65e$+00$ & 5.33e$+00$ & 6.82e$+00$ & 8.15e$+00$ & 1.09e$+01$ & 1.32e$+01$ \\
 12 & 14 & 1.29e$-01$ & 3.78e$-01$ & 1.02e$+00$ & 2.34e$+00$ & 3.50e$+00$ & 4.50e$+00$ & 5.36e$+00$ & 7.08e$+00$ & 8.39e$+00$ \\
 12 & 15 & 2.24e$-01$ & 1.03e$+00$ & 5.60e$+00$ & 3.08e$+01$ & 7.87e$+01$ & 1.48e$+02$ & 2.38e$+02$ & 5.39e$+02$ & 9.36e$+02$ \\
 12 & 16 & 1.03e$-01$ & 3.03e$-01$ & 8.47e$-01$ & 2.14e$+00$ & 3.51e$+00$ & 4.88e$+00$ & 6.19e$+00$ & 9.16e$+00$ & 1.17e$+01$ \\
 12 & 17 & 1.37e$-01$ & 6.24e$-01$ & 2.63e$+00$ & 1.12e$+01$ & 2.63e$+01$ & 4.80e$+01$ & 7.61e$+01$ & 1.73e$+02$ & 3.04e$+02$ \\
 12 & 18 & 2.53e$-02$ & 7.55e$-02$ & 1.86e$-01$ & 3.98e$-01$ & 5.80e$-01$ & 7.36e$-01$ & 8.71e$-01$ & 1.14e$+00$ & 1.34e$+00$ \\
 12 & 19 & 1.38e$-01$ & 3.76e$-01$ & 1.00e$+00$ & 2.77e$+00$ & 5.06e$+00$ & 7.77e$+00$ & 1.08e$+01$ & 1.97e$+01$ & 2.99e$+01$ \\
 12 & 20 & 2.73e$-01$ & 7.84e$-01$ & 1.98e$+00$ & 4.99e$+00$ & 9.29e$+00$ & 1.51e$+01$ & 2.25e$+01$ & 4.73e$+01$ & 7.95e$+01$ \\
 12 & 21 & 9.32e$-02$ & 2.52e$-01$ & 5.83e$-01$ & 1.19e$+00$ & 1.71e$+00$ & 2.17e$+00$ & 2.59e$+00$ & 3.46e$+00$ & 4.17e$+00$ \\
 12 & 22 & 6.65e$-02$ & 1.73e$-01$ & 3.76e$-01$ & 7.24e$-01$ & 1.02e$+00$ & 1.29e$+00$ & 1.53e$+00$ & 2.08e$+00$ & 2.54e$+00$ \\
 12 & 23 & 1.56e$-01$ & 4.62e$-01$ & 1.37e$+00$ & 4.40e$+00$ & 8.85e$+00$ & 1.44e$+01$ & 2.09e$+01$ & 3.97e$+01$ & 6.13e$+01$ \\
 12 & 24 & 7.48e$-02$ & 2.10e$-01$ & 4.97e$-01$ & 1.05e$+00$ & 1.58e$+00$ & 2.07e$+00$ & 2.54e$+00$ & 3.55e$+00$ & 4.38e$+00$ \\
 12 & 25 & 1.40e$-01$ & 3.93e$-01$ & 9.96e$-01$ & 2.42e$+00$ & 4.17e$+00$ & 6.29e$+00$ & 8.76e$+00$ & 1.63e$+01$ & 2.54e$+01$ \\
 12 & 26 & 1.91e$-02$ & 5.41e$-02$ & 1.33e$-01$ & 2.72e$-01$ & 3.81e$-01$ & 4.69e$-01$ & 5.43e$-01$ & 6.86e$-01$ & 7.90e$-01$ \\
 12 & 27 & 1.34e$-01$ & 4.08e$-01$ & 1.27e$+00$ & 3.80e$+00$ & 6.92e$+00$ & 1.04e$+01$ & 1.40e$+01$ & 2.38e$+01$ & 3.43e$+01$ \\
 12 & 28 & 4.70e$-02$ & 1.25e$-01$ & 3.28e$-01$ & 8.64e$-01$ & 1.52e$+00$ & 2.30e$+00$ & 3.18e$+00$ & 5.75e$+00$ & 8.64e$+00$ \\
 12 & 29 & 1.58e$-05$ & 9.00e$-05$ & 5.75e$-04$ & 3.81e$-03$ & 1.08e$-02$ & 2.17e$-02$ & 3.65e$-02$ & 8.69e$-02$ & 1.51e$-01$ \\
 12 & 30 & 4.26e$-02$ & 1.26e$-01$ & 3.47e$-01$ & 8.38e$-01$ & 1.30e$+00$ & 1.72e$+00$ & 2.10e$+00$ & 2.88e$+00$ & 3.51e$+00$ \\
 12 & 31 & 3.49e$-02$ & 9.35e$-02$ & 2.39e$-01$ & 5.71e$-01$ & 8.92e$-01$ & 1.19e$+00$ & 1.47e$+00$ & 2.08e$+00$ & 2.59e$+00$ \\
 12 & 32 & 3.73e$-02$ & 1.15e$-01$ & 3.23e$-01$ & 7.87e$-01$ & 1.23e$+00$ & 1.62e$+00$ & 1.97e$+00$ & 2.69e$+00$ & 3.24e$+00$ \\
 12 & 33 & 9.04e$-02$ & 3.07e$-01$ & 1.03e$+00$ & 3.33e$+00$ & 6.44e$+00$ & 1.01e$+01$ & 1.42e$+01$ & 2.54e$+01$ & 3.75e$+01$ \\
 12 & 34 & 3.47e$-02$ & 1.26e$-01$ & 4.52e$-01$ & 1.57e$+00$ & 3.18e$+00$ & 5.13e$+00$ & 7.34e$+00$ & 1.37e$+01$ & 2.08e$+01$ \\
 12 & 35 & 8.20e$-03$ & 2.04e$-02$ & 4.78e$-02$ & 1.18e$-01$ & 2.00e$-01$ & 2.85e$-01$ & 3.69e$-01$ & 5.67e$-01$ & 7.43e$-01$ \\
 12 & 36 & 1.14e$-02$ & 3.76e$-02$ & 1.46e$-01$ & 6.11e$-01$ & 1.34e$+00$ & 2.26e$+00$ & 3.31e$+00$ & 6.31e$+00$ & 9.61e$+00$ \\
 12 & 37 & 8.76e$-04$ & 3.68e$-03$ & 1.43e$-02$ & 4.96e$-02$ & 9.48e$-02$ & 1.46e$-01$ & 2.02e$-01$ & 3.58e$-01$ & 5.34e$-01$ \\
 13 & 15 & 1.71e$-01$ & 4.97e$-01$ & 1.35e$+00$ & 3.22e$+00$ & 5.04e$+00$ & 6.75e$+00$ & 8.32e$+00$ & 1.17e$+01$ & 1.44e$+01$ \\
 13 & 17 & 1.29e$-01$ & 3.40e$-01$ & 8.53e$-01$ & 1.86e$+00$ & 2.76e$+00$ & 3.54e$+00$ & 4.24e$+00$ & 5.67e$+00$ & 6.77e$+00$ \\
 13 & 19 & 4.08e$-02$ & 1.05e$-01$ & 2.44e$-01$ & 4.98e$-01$ & 7.12e$-01$ & 8.92e$-01$ & 1.05e$+00$ & 1.35e$+00$ & 1.58e$+00$ \\
 13 & 20 & 1.39e$-01$ & 3.51e$-01$ & 7.80e$-01$ & 1.56e$+00$ & 2.28e$+00$ & 2.93e$+00$ & 3.54e$+00$ & 4.84e$+00$ & 5.91e$+00$ \\
 13 & 23 & 1.57e$-01$ & 3.85e$-01$ & 8.36e$-01$ & 1.66e$+00$ & 2.40e$+00$ & 3.07e$+00$ & 3.66e$+00$ & 4.91e$+00$ & 5.89e$+00$ \\
 13 & 25 & 6.47e$-02$ & 1.75e$-01$ & 4.15e$-01$ & 8.48e$-01$ & 1.21e$+00$ & 1.53e$+00$ & 1.80e$+00$ & 2.37e$+00$ & 2.81e$+00$ \\
 13 & 27 & 5.77e$-02$ & 1.58e$-01$ & 4.04e$-01$ & 9.29e$-01$ & 1.42e$+00$ & 1.85e$+00$ & 2.24e$+00$ & 3.03e$+00$ & 3.64e$+00$ \\
 13 & 28 & 2.21e$-02$ & 6.05e$-02$ & 1.35e$-01$ & 2.59e$-01$ & 3.60e$-01$ & 4.45e$-01$ & 5.18e$-01$ & 6.64e$-01$ & 7.74e$-01$ \\
 13 & 33 & 4.96e$-02$ & 1.44e$-01$ & 3.84e$-01$ & 9.20e$-01$ & 1.44e$+00$ & 1.91e$+00$ & 2.33e$+00$ & 3.23e$+00$ & 3.94e$+00$ \\
 13 & 34 & 2.49e$-02$ & 8.19e$-02$ & 2.27e$-01$ & 5.47e$-01$ & 8.57e$-01$ & 1.15e$+00$ & 1.41e$+00$ & 1.97e$+00$ & 2.42e$+00$ \\
 13 & 36 & 8.59e$-03$ & 2.24e$-02$ & 5.27e$-02$ & 1.26e$-01$ & 2.10e$-01$ & 2.99e$-01$ & 3.87e$-01$ & 5.95e$-01$ & 7.80e$-01$ \\
 13 & 37 & 2.91e$-03$ & 1.26e$-02$ & 4.55e$-02$ & 1.31e$-01$ & 2.15e$-01$ & 2.91e$-01$ & 3.58e$-01$ & 4.91e$-01$ & 5.90e$-01$ \\
 14 & 15 & 6.88e$-02$ & 2.04e$-01$ & 5.68e$-01$ & 1.40e$+00$ & 2.23e$+00$ & 3.02e$+00$ & 3.76e$+00$ & 5.34e$+00$ & 6.62e$+00$ \\
 14 & 17 & 4.80e$-02$ & 1.29e$-01$ & 3.16e$-01$ & 6.82e$-01$ & 1.05e$+00$ & 1.42e$+00$ & 1.79e$+00$ & 2.64e$+00$ & 3.39e$+00$ \\
 14 & 19 & 2.18e$-02$ & 5.46e$-02$ & 1.21e$-01$ & 2.40e$-01$ & 3.45e$-01$ & 4.42e$-01$ & 5.34e$-01$ & 7.38e$-01$ & 9.12e$-01$ \\
 14 & 20 & 7.73e$-02$ & 1.99e$-01$ & 4.42e$-01$ & 8.86e$-01$ & 1.29e$+00$ & 1.67e$+00$ & 2.01e$+00$ & 2.73e$+00$ & 3.30e$+00$ \\
 14 & 23 & 6.50e$-02$ & 1.83e$-01$ & 4.52e$-01$ & 9.83e$-01$ & 1.48e$+00$ & 1.92e$+00$ & 2.33e$+00$ & 3.17e$+00$ & 3.83e$+00$ \\
 14 & 25 & 2.99e$-02$ & 8.21e$-02$ & 1.92e$-01$ & 3.94e$-01$ & 5.71e$-01$ & 7.29e$-01$ & 8.74e$-01$ & 1.19e$+00$ & 1.46e$+00$ \\
 14 & 27 & 3.38e$-02$ & 9.54e$-02$ & 2.42e$-01$ & 5.45e$-01$ & 8.28e$-01$ & 1.08e$+00$ & 1.31e$+00$ & 1.78e$+00$ & 2.13e$+00$ \\
 14 & 28 & 1.21e$-02$ & 2.92e$-02$ & 6.51e$-02$ & 1.32e$-01$ & 1.92e$-01$ & 2.44e$-01$ & 2.92e$-01$ & 3.93e$-01$ & 4.76e$-01$ \\
 14 & 33 & 3.02e$-02$ & 8.52e$-02$ & 2.17e$-01$ & 4.96e$-01$ & 7.65e$-01$ & 1.01e$+00$ & 1.24e$+00$ & 1.71e$+00$ & 2.08e$+00$ \\
 14 & 34 & 1.56e$-02$ & 4.49e$-02$ & 1.19e$-01$ & 2.89e$-01$ & 4.57e$-01$ & 6.17e$-01$ & 7.68e$-01$ & 1.11e$+00$ & 1.41e$+00$ \\
 14 & 36 & 4.58e$-03$ & 1.38e$-02$ & 3.81e$-02$ & 9.49e$-02$ & 1.57e$-01$ & 2.21e$-01$ & 2.86e$-01$ & 4.45e$-01$ & 5.90e$-01$ \\
 14 & 37 & 3.69e$-03$ & 1.41e$-02$ & 4.32e$-02$ & 1.05e$-01$ & 1.58e$-01$ & 2.02e$-01$ & 2.37e$-01$ & 3.06e$-01$ & 3.55e$-01$ \\
 15 & 16 & 1.21e$-01$ & 4.23e$-01$ & 1.23e$+00$ & 2.96e$+00$ & 4.70e$+00$ & 6.45e$+00$ & 8.18e$+00$ & 1.23e$+01$ & 1.61e$+01$ \\
 15 & 17 & 5.59e$-01$ & 1.65e$+00$ & 4.40e$+00$ & 1.13e$+01$ & 2.00e$+01$ & 3.06e$+01$ & 4.32e$+01$ & 8.55e$+01$ & 1.48e$+02$ \\
 15 & 18 & 3.91e$-02$ & 1.13e$-01$ & 2.54e$-01$ & 4.74e$-01$ & 6.43e$-01$ & 7.84e$-01$ & 9.05e$-01$ & 1.15e$+00$ & 1.34e$+00$ \\
 15 & 19 & 1.58e$-01$ & 4.01e$-01$ & 9.20e$-01$ & 2.13e$+00$ & 3.61e$+00$ & 5.31e$+00$ & 7.20e$+00$ & 1.25e$+01$ & 1.83e$+01$ \\
 15 & 20 & 4.14e$-01$ & 1.19e$+00$ & 3.20e$+00$ & 9.53e$+00$ & 1.95e$+01$ & 3.32e$+01$ & 5.04e$+01$ & 1.07e$+02$ & 1.82e$+02$ \\
 15 & 21 & 1.50e$-01$ & 3.63e$-01$ & 7.71e$-01$ & 1.47e$+00$ & 2.07e$+00$ & 2.61e$+00$ & 3.11e$+00$ & 4.23e$+00$ & 5.21e$+00$ \\
 15 & 22 & 1.10e$-01$ & 2.78e$-01$ & 5.85e$-01$ & 1.06e$+00$ & 1.44e$+00$ & 1.79e$+00$ & 2.11e$+00$ & 2.83e$+00$ & 3.43e$+00$ \\
 15 & 23 & 1.91e$-01$ & 5.87e$-01$ & 1.73e$+00$ & 5.42e$+00$ & 1.10e$+01$ & 1.87e$+01$ & 2.84e$+01$ & 6.13e$+01$ & 1.05e$+02$ \\
 15 & 24 & 1.51e$-01$ & 3.55e$-01$ & 7.42e$-01$ & 1.43e$+00$ & 2.07e$+00$ & 2.69e$+00$ & 3.28e$+00$ & 4.67e$+00$ & 5.91e$+00$ \\
 15 & 25 & 1.75e$-01$ & 5.16e$-01$ & 1.35e$+00$ & 3.42e$+00$ & 5.95e$+00$ & 8.86e$+00$ & 1.21e$+01$ & 2.12e$+01$ & 3.12e$+01$ \\
 15 & 26 & 3.34e$-02$ & 8.43e$-02$ & 1.80e$-01$ & 3.28e$-01$ & 4.38e$-01$ & 5.27e$-01$ & 6.03e$-01$ & 7.52e$-01$ & 8.67e$-01$ \\
 15 & 27 & 1.78e$-01$ & 5.45e$-01$ & 1.59e$+00$ & 4.68e$+00$ & 8.83e$+00$ & 1.37e$+01$ & 1.90e$+01$ & 3.35e$+01$ & 4.89e$+01$ \\
 15 & 28 & 7.03e$-02$ & 1.99e$-01$ & 5.01e$-01$ & 1.15e$+00$ & 1.85e$+00$ & 2.61e$+00$ & 3.39e$+00$ & 5.39e$+00$ & 7.34e$+00$ \\
 15 & 29 & 3.23e$-06$ & 2.35e$-05$ & 1.38e$-04$ & 7.75e$-04$ & 2.24e$-03$ & 4.60e$-03$ & 7.73e$-03$ & 1.77e$-02$ & 2.91e$-02$ \\
 15 & 30 & 4.89e$-02$ & 1.43e$-01$ & 4.02e$-01$ & 1.00e$+00$ & 1.60e$+00$ & 2.17e$+00$ & 2.71e$+00$ & 3.89e$+00$ & 4.88e$+00$ \\
 15 & 31 & 3.48e$-02$ & 9.97e$-02$ & 2.67e$-01$ & 6.51e$-01$ & 1.05e$+00$ & 1.45e$+00$ & 1.84e$+00$ & 2.71e$+00$ & 3.43e$+00$ \\
 15 & 32 & 5.55e$-02$ & 1.52e$-01$ & 3.92e$-01$ & 9.24e$-01$ & 1.44e$+00$ & 1.91e$+00$ & 2.34e$+00$ & 3.28e$+00$ & 4.05e$+00$ \\
 15 & 33 & 1.16e$-01$ & 4.02e$-01$ & 1.27e$+00$ & 3.84e$+00$ & 7.46e$+00$ & 1.21e$+01$ & 1.78e$+01$ & 3.58e$+01$ & 5.82e$+01$ \\
 15 & 34 & 6.66e$-02$ & 2.53e$-01$ & 8.83e$-01$ & 2.77e$+00$ & 5.12e$+00$ & 7.71e$+00$ & 1.05e$+01$ & 1.79e$+01$ & 2.58e$+01$ \\
 15 & 35 & 9.74e$-03$ & 2.71e$-02$ & 7.63e$-02$ & 2.21e$-01$ & 3.85e$-01$ & 5.46e$-01$ & 6.98e$-01$ & 1.03e$+00$ & 1.32e$+00$ \\
 15 & 36 & 1.96e$-02$ & 7.53e$-02$ & 3.25e$-01$ & 1.26e$+00$ & 2.53e$+00$ & 4.04e$+00$ & 5.73e$+00$ & 1.06e$+01$ & 1.59e$+01$ \\
 15 & 37 & 3.71e$-03$ & 1.74e$-02$ & 6.28e$-02$ & 1.73e$-01$ & 2.82e$-01$ & 3.85e$-01$ & 4.82e$-01$ & 7.05e$-01$ & 9.08e$-01$ \\
 16 & 17 & 6.39e$-02$ & 2.12e$-01$ & 6.15e$-01$ & 1.46e$+00$ & 2.23e$+00$ & 2.93e$+00$ & 3.55e$+00$ & 4.86e$+00$ & 5.90e$+00$ \\
 16 & 19 & 3.93e$-02$ & 1.05e$-01$ & 2.45e$-01$ & 4.87e$-01$ & 6.80e$-01$ & 8.43e$-01$ & 9.84e$-01$ & 1.27e$+00$ & 1.50e$+00$ \\
 16 & 20 & 1.66e$-01$ & 4.26e$-01$ & 9.54e$-01$ & 1.86e$+00$ & 2.63e$+00$ & 3.31e$+00$ & 3.93e$+00$ & 5.30e$+00$ & 6.47e$+00$ \\
 16 & 23 & 1.53e$-01$ & 3.68e$-01$ & 8.15e$-01$ & 1.65e$+00$ & 2.42e$+00$ & 3.16e$+00$ & 3.85e$+00$ & 5.45e$+00$ & 6.86e$+00$ \\
 16 & 25 & 7.57e$-02$ & 2.09e$-01$ & 4.90e$-01$ & 9.57e$-01$ & 1.33e$+00$ & 1.65e$+00$ & 1.94e$+00$ & 2.54e$+00$ & 3.03e$+00$ \\
 16 & 27 & 7.05e$-02$ & 2.07e$-01$ & 5.46e$-01$ & 1.26e$+00$ & 1.94e$+00$ & 2.57e$+00$ & 3.15e$+00$ & 4.40e$+00$ & 5.41e$+00$ \\
 16 & 28 & 2.73e$-02$ & 6.95e$-02$ & 1.55e$-01$ & 3.03e$-01$ & 4.21e$-01$ & 5.19e$-01$ & 6.03e$-01$ & 7.73e$-01$ & 9.04e$-01$ \\
 16 & 33 & 7.84e$-02$ & 2.09e$-01$ & 5.09e$-01$ & 1.17e$+00$ & 1.83e$+00$ & 2.46e$+00$ & 3.07e$+00$ & 4.44e$+00$ & 5.62e$+00$ \\
 16 & 34 & 3.25e$-02$ & 1.01e$-01$ & 2.73e$-01$ & 6.78e$-01$ & 1.11e$+00$ & 1.54e$+00$ & 1.95e$+00$ & 2.84e$+00$ & 3.56e$+00$ \\
 16 & 36 & 8.10e$-03$ & 2.43e$-02$ & 7.04e$-02$ & 2.09e$-01$ & 3.76e$-01$ & 5.50e$-01$ & 7.23e$-01$ & 1.12e$+00$ & 1.48e$+00$ \\
 16 & 37 & 6.15e$-03$ & 2.99e$-02$ & 1.11e$-01$ & 3.00e$-01$ & 4.74e$-01$ & 6.25e$-01$ & 7.54e$-01$ & 1.01e$+00$ & 1.19e$+00$ \\
 17 & 18 & 1.77e$-02$ & 5.49e$-02$ & 1.40e$-01$ & 2.94e$-01$ & 4.29e$-01$ & 5.50e$-01$ & 6.60e$-01$ & 8.94e$-01$ & 1.08e$+00$ \\
 17 & 19 & 1.36e$-01$ & 5.67e$-01$ & 2.27e$+00$ & 9.74e$+00$ & 2.33e$+01$ & 4.30e$+01$ & 6.89e$+01$ & 1.60e$+02$ & 2.87e$+02$ \\
 17 & 20 & 3.48e$-01$ & 1.19e$+00$ & 3.93e$+00$ & 1.39e$+01$ & 3.19e$+01$ & 6.02e$+01$ & 9.99e$+01$ & 2.53e$+02$ & 4.84e$+02$ \\
 17 & 21 & 1.34e$-01$ & 3.42e$-01$ & 7.31e$-01$ & 1.32e$+00$ & 1.76e$+00$ & 2.10e$+00$ & 2.38e$+00$ & 2.93e$+00$ & 3.33e$+00$ \\
 17 & 22 & 7.91e$-02$ & 2.03e$-01$ & 4.27e$-01$ & 7.75e$-01$ & 1.05e$+00$ & 1.28e$+00$ & 1.48e$+00$ & 1.90e$+00$ & 2.23e$+00$ \\
 17 & 23 & 2.92e$-01$ & 1.04e$+00$ & 3.55e$+00$ & 1.35e$+01$ & 3.03e$+01$ & 5.32e$+01$ & 8.12e$+01$ & 1.69e$+02$ & 2.75e$+02$ \\
 17 & 24 & 1.14e$-01$ & 2.99e$-01$ & 6.57e$-01$ & 1.26e$+00$ & 1.77e$+00$ & 2.21e$+00$ & 2.60e$+00$ & 3.37e$+00$ & 3.95e$+00$ \\
 17 & 25 & 1.87e$-01$ & 6.64e$-01$ & 2.05e$+00$ & 6.11e$+00$ & 1.19e$+01$ & 1.94e$+01$ & 2.88e$+01$ & 5.90e$+01$ & 9.77e$+01$ \\
 17 & 26 & 2.12e$-02$ & 6.11e$-02$ & 1.42e$-01$ & 2.67e$-01$ & 3.58e$-01$ & 4.29e$-01$ & 4.87e$-01$ & 5.98e$-01$ & 6.79e$-01$ \\
 17 & 27 & 1.31e$-01$ & 4.62e$-01$ & 1.52e$+00$ & 4.70e$+00$ & 8.93e$+00$ & 1.40e$+01$ & 1.97e$+01$ & 3.64e$+01$ & 5.58e$+01$ \\
 17 & 28 & 4.91e$-02$ & 1.62e$-01$ & 4.85e$-01$ & 1.38e$+00$ & 2.50e$+00$ & 3.88e$+00$ & 5.58e$+00$ & 1.13e$+01$ & 1.92e$+01$ \\
 17 & 29 & 3.10e$-06$ & 1.66e$-05$ & 9.34e$-05$ & 6.09e$-04$ & 1.64e$-03$ & 3.11e$-03$ & 4.95e$-03$ & 1.09e$-02$ & 1.84e$-02$ \\
 17 & 30 & 3.81e$-02$ & 1.05e$-01$ & 2.71e$-01$ & 6.07e$-01$ & 9.02e$-01$ & 1.16e$+00$ & 1.37e$+00$ & 1.80e$+00$ & 2.12e$+00$ \\
 17 & 31 & 2.29e$-02$ & 6.50e$-02$ & 1.64e$-01$ & 3.72e$-01$ & 5.71e$-01$ & 7.55e$-01$ & 9.23e$-01$ & 1.29e$+00$ & 1.58e$+00$ \\
 17 & 32 & 3.29e$-02$ & 9.61e$-02$ & 2.50e$-01$ & 5.78e$-01$ & 8.83e$-01$ & 1.15e$+00$ & 1.38e$+00$ & 1.84e$+00$ & 2.18e$+00$ \\
 17 & 33 & 1.02e$-01$ & 4.17e$-01$ & 1.64e$+00$ & 5.79e$+00$ & 1.15e$+01$ & 1.81e$+01$ & 2.54e$+01$ & 4.53e$+01$ & 6.64e$+01$ \\
 17 & 34 & 4.22e$-02$ & 1.85e$-01$ & 8.22e$-01$ & 3.38e$+00$ & 7.34e$+00$ & 1.24e$+01$ & 1.82e$+01$ & 3.55e$+01$ & 5.51e$+01$ \\
 17 & 35 & 8.38e$-03$ & 2.18e$-02$ & 5.03e$-02$ & 1.17e$-01$ & 1.88e$-01$ & 2.60e$-01$ & 3.28e$-01$ & 4.80e$-01$ & 6.06e$-01$ \\
 17 & 36 & 1.15e$-02$ & 4.83e$-02$ & 1.75e$-01$ & 6.24e$-01$ & 1.36e$+00$ & 2.37e$+00$ & 3.63e$+00$ & 7.74e$+00$ & 1.30e$+01$ \\
 17 & 37 & 1.77e$-03$ & 6.31e$-03$ & 1.87e$-02$ & 4.91e$-02$ & 8.50e$-02$ & 1.28e$-01$ & 1.79e$-01$ & 3.41e$-01$ & 5.50e$-01$ \\
 \end{longtable}
 
  
\end{document}